\documentclass[%
 reprint,
 longbibliography,
 amsmath,amssymb,
 aps,prx,
]{revtex4-2}
\usepackage{txfonts}
\usepackage{siunitx}
\usepackage{graphicx}
\usepackage{dcolumn}
\usepackage{bm}
\usepackage[colorlinks,allcolors= blue]{hyperref}
\begin{document}

\newcount\colveccount
\newcommand*\colvec[1]{
        \global\colveccount#1
        \begin{pmatrix}
        \colvecnext
}
\def\colvecnext#1{
        #1
        \global\advance\colveccount-1
        \ifnum\colveccount>0
                \\
                \expandafter\colvecnext
        \else
                \end{pmatrix}
        \fi
}

\title{Optical coherent feedback control of a mechanical oscillator}

\author{Maryse Ernzer}\altaffiliation{These authors contributed equally to this work.}
\author{Manel Bosch Aguilera}
\altaffiliation{These authors contributed equally to this work.}
\author{Matteo Brunelli}
\author{Gian-Luca Schmid}
\author{{Thomas M. Karg}}
\altaffiliation{Present address: IBM Research Europe, Zurich, S\"aumerstrasse 4, 8803 R\"uschlikon, Switzerland.}
\author{Christoph Bruder}
\author{Patrick P. Potts}
\email{patrick.potts@unibas.ch}
\author{Philipp Treutlein}
\email{philipp.treutlein@unibas.ch}
\affiliation{Department of Physics and Swiss Nanoscience Institute,
University of Basel, Klingelbergstrasse 82, 4056 Basel, Switzerland}%

\begin{abstract}
Feedback is a powerful and ubiquitous technique both in classical and quantum system control. Its standard implementation relies on measuring the state of a system, processing {the classical signal,} and feeding {it} back {to the system}. In quantum physics, however, measurements not only read out the state of the system but also modify it irreversibly. Coherent feedback is a different kind of feedback that coherently processes and feeds back quantum signals without actually measuring the system. 
Here, we report on the {experimental} realization {and the theoretical analysis} of an optical coherent feedback platform to control the motional state of a nanomechanical membrane in an optical cavity. The coherent feedback loop consists of a light field interacting twice with the same mechanical mode through different cavity modes, without {performing any} measurement. Tuning the optical phase and delay of the feedback loop allows us to control the motional state of the mechanical oscillator, its resonance frequency and {also its} damping rate, {which we use} to cool the membrane close to the quantum ground state. Our theoretical analysis provides the optimal cooling conditions, showing that this new technique enables ground-state cooling. Experimentally, we show that we can cool the membrane to a state with $\bar{n}_m = 4.89 \pm  0.14 $ phonons (\SI{480}{\micro K}) in a $\SI{20}{K}$ environment. This lies below the theoretical limit of cavity dynamical backaction cooling in the unresolved sideband regime {and is achieved with only 1$\%$ of the optical power required for cavity cooling}. Our feedback scheme is very versatile, offering new opportunities for quantum control in a variety of optomechanical systems.  

\end{abstract}

\maketitle

\begin{figure*}[thb]
    \centering
    \includegraphics[width=\textwidth]{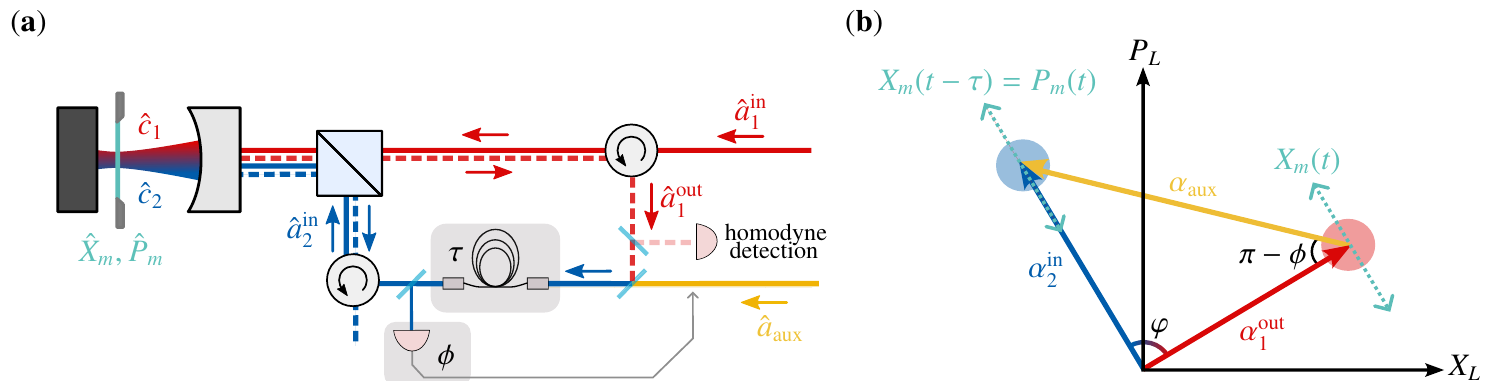}
    \caption{Sketch of the experimental setup and optical coherent feedback loop. \textbf{(a):} An incoming light beam $\hat{a}^\mathrm{in}_1$ is injected into an optomechanical cavity, where it drives the cavity field $\hat{c}_1$ that interacts with a mechanical oscillator with quadratures $\hat
    X_m,\hat{P}_m$. The back-reflected beam $\hat{a}^\mathrm{out}_1$ is combined with an auxiliary local oscillator mode $\hat{a}_{\mathrm{aux}}$ to control the phase of the feedback loop $\varphi$. The combined field is delayed by $\tau$ with the help of an optical fiber, before being sent back as input $\hat{a}^\mathrm{in}_2$ for a second interaction with the mechanical oscillator in an orthogonal polarization cavity mode $\hat{c}_2$. The outgoing light after the second interaction leaves the loop. A small fraction of $\hat{a}^\mathrm{out}_1$ is picked up for detection and phase locking of the loop. {The color coding of the light beams is used for visual guidance. Dashed lines are used for visual distinction between the incoming and back-reflected beams. } \textbf{(b):} Phase-space visualisation of the feedback loop. The sketch represents the amplitude $(X_L)$ and phase $(P_L)$ quadratures of the light outside the cavity in an arbitrary frame. On resonance, the coherent cavity output field after the first interaction  $\alpha_1^\mathrm{out}$ is phase-modulated (dashed line) with the membrane displacement signal $X_m(t)$. This is converted into an amplitude modulation of $\alpha_2^\mathrm{in}$ by mixing the coherent field with an auxiliary local oscillator $\alpha_\mathrm{aux}$ with the appropriate relative phase $\phi$, {to achieve the desired value of $\varphi$ between $\alpha_1$ and $\alpha_2$}. 
    After a delay $\Omega_m\tau = \pi/2$, the amplitude modulation becomes proportional to the momentum of the membrane  $P_m$ at time $t$ (dashed line in  $\alpha_2^\mathrm{in}$) and exerts a force on the mechanical oscillator.}
    \label{fig:SetupPhaseSketch}
\end{figure*}

\section{Introduction}

Quantum feedback is a powerful technique for cooling and controlling quantum systems \cite{zhang2017}. The conventional strategy relies on quantum-limited measurements followed by classical processing and feedback actuation onto the system. However, quantum mechanics also allows for {\emph{coherent}} feedback of {\emph{quantum}} signals \cite{lloyd2000,wiseman1994}, without destroying coherence in the process. 
This kind of feedback may exploit the information stored in non-commuting observables while circumventing the decoherence and back-action noise associated with a measurement \cite{lloyd2000, wiseman2009,zhang2017}. 
Coherent feedback has thus the potential to improve quantum control and provide new capabilities {across} a broad range of physical systems \cite{yamamoto2014, jacobs2014}. Coherent feedback strategies have {so far} been adopted to assist in a variety of different tasks \cite{zhang2017}, e.g.~for noise cancellation \cite{mabuchi2008, tsang2010}, pure-state preparation \cite{jacobs2014},  optical squeezing \cite{gough2009,iida2012}, stabilization and enhancement of entanglement  \cite{shankar2013,zhou2015}, sympathetic cooling \cite{rohde2001,frimmer2016,schmid2022a}, swaps of arbitrary states \cite{nelson2000}, qubit state control \cite{hirose2016a}, and generating large optical nonlinearities at the single-photon level \cite{zhang2012,wang2017a}.

Optomechanical systems are very well suited for coherent feedback control, as they offer a clean and tailored interface between highly coherent mechanical and electromagnetic field modes \cite{aspelmeyer2014}. Indeed, various coherent feedback protocols have been theoretically proposed to enhance the cooling of optomechanical systems \cite{huang2019, guo2022, mansouri2022}, to reduce the added noise in the low phonon-number regime of optomechanical precision measurements \cite{hamerly2012}, to enable or enhance entanglement generation, verification, as well as state transfer \cite{ li2017, amazioug2020, harwood2021, guo2022}. 
Coherent feedback {can thus facilitate and extend} the capabilities of quantum transducers between optics and mechanics \cite{barzanjeh2022}. 

Despite this wide range of possibilities, there have been surprisingly few experiments investigating coherent feedback in optomechanics \cite{schmid2022a, kerckhoff2013}. An optical coherent feedback loop acting directly onto a mechanical oscillator has not yet been realized. 
    Moreover, while measurement-based feedback has been studied in some depth from a theory point of view \cite{mancini1998, rossi2018, delic2020, tebbenjohanns2021, whittle2021}, essential questions regarding the performance {and limitations of coherent feedback in actual optomechanics  experiments} remain open. 

In this work, we present both a theoretical description and experimental realization of a simple, all-optical coherent feedback platform to control a single vibrational mode of a mechanical oscillator. We use a double-pass scheme where an optical signal interacts twice with the same mechanical mode through two different cavity modes of orthogonal polarization. The entire control of the phase and delay of the feedback signal is implemented purely via the optical field, without introducing measurements and subsequent electronic processing. Our approach is thus able to generate a variety of different interactions, ranging from Hamiltonian couplings to dissipative and non-reciprocal dynamics \cite{karg2019, metelmann2015, metelmann2017}.

As a first application of the extended control offered by the coherent feedback loop, we investigate {the} cooling of the mechanical mode close to its quantum ground state, {which is} a prerequisite for many applications in quantum science and technology \cite{aspelmeyer2014, chu2020, monsel2021,barzanjeh2022}.  Theoretically, we show that coherent feedback enables ground-state cooling even in the {unresolved} sideband regime, where cavity dynamical backaction cooling with a single or two independent optomechanical interactions cannot reach the ground state \cite{aspelmeyer2014}. Experimentally, we demonstrate the advantage of the coherent feedback loop by cooling below the theoretical limit of cavity dynamical backaction cooling in our system. 
This is particularly interesting for optomechanical systems with cavities of large bandwidth, which induce only a small delay and 
are frequently encountered in optomechanical displacement sensing, quantum interfaces and hybrid setups \cite{treutlein2014}.

The remainder of this paper is structured as follows: We first provide an overview of the working principle of our coherent feedback platform for controlling a mechanical oscillator in an optical cavity. Next, we develop a theoretical model of the feedback scheme, followed by our experimental results on motional state control and its application to cooling. Finally, we compare our theoretical results with those of measurement-based feedback for the specific task of cooling.

\section{Overview of the coherent feedback scheme}
\label{sec:overview}
We start by illustrating the working principle of our coherent feedback scheme, sketched in Fig.~\ref{fig:SetupPhaseSketch}. The goal is to control the motional state of a mechanical oscillator by designing an optical feedback loop that preserves the quantum coherent properties of the light field, which acts as the controller. To this end, the mechanical oscillator is radiation-pressure coupled to \emph{two} cavity modes in a cascaded double-pass interaction. The first interaction takes place between the mechanical oscillator and the cavity mode $\hat c_1$, which is driven by a strong local oscillator, realizing the standard cavity optomechanical interaction \cite{aspelmeyer2014}. 
Due to the optomechanical coupling, information about the mechanical position $\hat X_m$ is imprinted onto the phase quadrature of $\hat c_1$. This mode is then cascaded into the second cavity mode $\hat c_2$ via an all-optical feedback loop. Specifically, the output light of the first mode, with mean amplitude $\alpha_1^\mathrm{out}$, is mixed with a second local oscillator and fed back as the input of the second cavity mode, with amplitude $\alpha_2^\mathrm{in}$, as shown in Fig.~\ref{fig:SetupPhaseSketch}. 
The resulting optical feedback loop is characterized by two parameters, the relative phase $\varphi$ and the in-loop delay time $\tau$. The phase $\varphi$ is {controlled} by the second local oscillator $\alpha_\mathrm{aux}$, which implements a displacement  in the optical phase space of the modes traveling within the loop, see Fig.~\ref{fig:SetupPhaseSketch}\,(b).

Both feedback parameters are crucial for controlling the mechanical oscillator. The phase $\varphi$ is adjusted so that the phase quadrature of the outgoing mode, which contains information on the mechanical position, is turned into the amplitude quadrature of the incoming mode, {such that it} exerts {a feedback force by radiation pressure} on the mechanical oscillator. As sketched in Fig.~\ref{fig:SetupPhaseSketch}\, (b), this occurs for $\varphi=\pi/2$. Adjusting the delay $\tau$ allows to either feed back the instantaneous position [when $\hat X_m(t-\tau)\simeq\hat X_m(t)$], momentum [when $\hat X_m(t-\tau)\simeq\hat P_m(t)$, as represented in Fig.~\ref{fig:SetupPhaseSketch}~(b)], or a superposition thereof. While feeding back the position enables control of the mechanical oscillator frequency, feeding back the momentum allows control of its damping, which can be exploited for ground-state cooling.

Previous theoretical proposals for coherent feedback cooling of mechanical oscillators \cite{huang2019, harwood2021, guo2022, mansouri2022} rely on coherently enhancing the interaction of the cavity light with the mechanics, mostly by  modifying the effective cavity linewidth \cite{harwood2021, guo2022}, and on loops that impart only a delay (plus unavoidable coupling losses). 
In contrast, our scheme applies the coherent feedback directly to the mechanical oscillator, such that the feedback can be generated with a single cavity driven in two independent modes. Moreover, it allows tuning of the loop phase $\varphi$, which strongly influences the effect of the feedback.  

Our scheme requires no additional optical devices such as cavities and only minor modifications of the optical path, resulting in a modular scheme that is optimally suited for incorporation into various types of optomechanical systems. 

Furthermore, our double-pass scheme does not require non-classical input light states \cite{clark2017, schafermeier2016,monsel2021}, additional interactions with other physical systems \cite{rohde2001, christoph2018, schmid2022a}, nor the overall very high {detector} efficiency of measurement-based feedback schemes \cite{mancini1998, rossi2018, delic2020, tebbenjohanns2021, whittle2021}, {which is now replaced by the requirement of small optical losses in the loop}.
The relaxation of the requirements on measurement efficiency renders our scheme valuable for systems working in wavelength ranges where efficient photodetectors are not available, e.g. in integrated circuit platforms \cite{rogalski2017}.

\section{Theoretical model}

In this section, we provide a theoretical model for the coherent feedback scheme described above.

{\subsection{Langevin equations}}
The mechanical oscillator that is to be controlled is described by the {linearized} Langevin equations 
\begin{equation}
\label{eq:langevinmech}
\begin{aligned}
&\partial_t\hat{X}_m(t) = \Omega_m\hat{P}_m(t),\\&\partial_t\hat{P}_m(t) = -\Omega_m\hat{X}_m(t)-\gamma_m\hat{P}_m(t)-2\sum_{j=1}^2g_j\hat{x}_j(t)-\sqrt{2}\,\hat{\xi}_{\rm th}(t),\\
\end{aligned}
\end{equation}
where $\hat{X}_m(t)$ and $\hat{P}_m(t)$ denote the  dimensionless position and momentum operators of the mechanical oscillator, $\Omega_m$ its frequency, and $\gamma_m$ its  energy damping rate. The mechanical oscillator is driven by thermal noise $\hat{\xi}_{\rm th}$, which has zero average and is fully described by its spectral density \cite{giovannetti2001}
{
\begin{equation}
    \label{eq:thermalspec}
    S_{\rm th}(\omega) = \gamma_m\frac{|\omega|}{\Omega_m}[n_{B}(|\omega|)+\Theta(\omega)], 
\end{equation}}
where $n_{B}(\omega) = [\exp(\hbar\omega/k_BT)-1]^{-1}$ denotes the Bose-Einstein distribution and $\Theta(\omega)$ the Heaviside step function.

The oscillator furthermore couples to the amplitude quadrature of two optical modes, $\hat{x}_j = (\hat{c}_j^\dagger+\hat{c}_j)/\sqrt{2}$ with strength $g_j$, where $j=1,2$. As discussed in more detail below, the coupling strengths depend on the average displacements of the optical modes and the operators $\hat{c}_j$ describe fluctuations around these displacements \cite{aspelmeyer2014}. 

The first cavity mode is described by the Langevin equation (in a frame rotating at the laser frequency $\omega_{L}$)
\begin{equation}
\label{eq:langevinc1}
\partial_t\hat{c}_1(t) = \left(i\Delta-\frac{\kappa}{2}\right)\hat{c}_1(t)-i\sqrt{2}g_1\hat{X}_m(t)-\sqrt{\kappa}\hat{a}_{1}^{\rm in}(t),
\end{equation}
where $\Delta=\omega_{L}-\omega_c$ denotes the detuning from the cavity mode frequency $\omega_c$ and $\kappa$ the cavity linewidth. This mode is driven by a local oscillator with frequency $\omega_{L}$ and average displacement $\alpha_1^{\rm in}$. 
Fluctuations around this displacement are described by the operator $\hat{a}_{1}^{\rm in}$. Due to the optomechanical coupling, the outgoing field leaving this cavity mode
\begin{equation}
    \label{eq:a1out}
    \hat{a}_1^{\rm out}(t) = \hat{a}_1^{\rm in}(t)+\sqrt{\kappa}\hat{c}_1(t),
\end{equation}
contains information on the position of the mechanical oscillator. From Eq.~\eqref{eq:langevinc1}, it follows that on resonance ($\Delta=0$), this information is only contained in the phase quadrature $\hat{p}_1 = i(\hat{c}_1^\dagger-\hat{c}_1)/\sqrt{2}$.

To implement the coherent feedback, the output of the first cavity mode is fed back into the input of the second cavity mode. Before it is coupled into the cavity, it undergoes a displacement by combining it with an auxiliary local oscillator with the same frequency $\omega_{L}$ and average displacement $\alpha_{\rm aux}$, and it is delayed by the time $\tau$. The second cavity mode is then driven by the input mode \cite{gardiner2000}
\begin{equation}
    \label{eq:a2in}
    \hat{a}_2^{\rm in} (t) = \sqrt{\eta}\,e^{i\varphi}\hat{a}_1^{\rm out}(t-\tau)+\sqrt{1-\eta}\,\hat{a}_{\rm aux}(t),
\end{equation}
where the phase $\varphi = {\rm arg}(\alpha_1/\alpha_2)$ denotes the phase difference between the average displacements of the cavity modes. As shown in App.~\ref{ap:langevin}, $\eta$ takes into account any losses in the loop. {We note that we chose to define the mode $\hat{a}_{\rm aux}$ such that no time-shift appears in its argument.}
The Langevin equation for the second cavity mode then reads
\begin{equation}
\label{eq:langevinc2}
\begin{aligned}
&\partial_t\hat{c}_2(t) = \left(i\Delta-\frac{\kappa}{2}\right)\hat{c}_2(t)-i\sqrt{2}g_2\hat{X}_m(t)-\sqrt{\eta}\kappa \,e^{i\varphi}\hat{c}_1(t-\tau)\\&\hspace{1cm}-\sqrt{\eta\kappa}\,e^{i\varphi}\hat{a}_1^{\rm in}(t-\tau)-\sqrt{(1-\eta)\kappa}\,\hat{a}_{\rm aux}(t).
\end{aligned}
\end{equation}
{The average displacements $\alpha_j$ of the cavity modes are given in Eq.~\eqref{eq:means}.}
In terms of these, the optomechanical coupling strengths can be written as $g_j = g_0|\alpha_j|$, where $g_0$ is the bare coupling strength. The amplitudes of the local oscillators are related to their input powers as $P_1 = \hbar\omega_{L}|\alpha_1^{\rm in}|^2$ and $P_{\rm aux} = \hbar\omega_{L}|\alpha_{\rm aux}|^2$ respectively.

To better understand the effect of the coherent feedback loop, it is illustrative to eliminate the cavity modes from the Langevin equations. For a high-quality oscillator $\gamma_m\ll\Omega_m$ and delay times that obey $\tau\gamma_m\ll 1$, we find
\begin{equation}
\label{eq:dampharmx}
\begin{aligned}
\partial_t^2\hat{X}_m(t) =& -(\Omega_m+\delta\Omega_m)^2\hat{X}_m(t)-(\gamma_m+\Gamma_m)\,\partial_t\hat{X}_m(t)\\&+\Omega_m\sqrt{2}\,\hat{\xi}_{\rm th}(t)+\Omega_m\sqrt{2}\,\hat{\xi}_{\rm fb}(t),
\end{aligned}
\end{equation}
where $\Gamma_m$ and $\delta\Omega_m$ denote the effective damping and the frequency shift that are controlled by the coherent feedback loop. {To derive Eq.~\eqref{eq:dampharmx}, we assumed that these are small compared to the frequency of the resonator, i.e., $\Gamma_m, \delta\Omega_m\ll \Omega_m$. In the unresolved sideband regime and on cavity resonance, we find}
\begin{equation}
    \label{eq:dampshift}
    \begin{aligned}
    &\Gamma_m = 16\sqrt{\eta}\frac{g_1g_2}{\kappa}\sin(\varphi)\sin(\Omega_m\tau),\\
    &\delta \Omega_m =-8\sqrt{\eta}\frac{g_1g_2}{\kappa}\sin(\varphi)\cos(\Omega_m\tau).
    \end{aligned}
\end{equation}
Expressions for the general scenario are given in App.~\ref{ap:theo}. 

From these expressions, we may understand the physical significance of the parameters that determine the coherent feedback loop. To maximize the effect that the optical field exerts on the mechanics, we should choose $\varphi=\pi/2$. The reason for this is sketched in Fig.~\ref{fig:SetupPhaseSketch}\,(b) and was already discussed qualitatively in Sec.~\ref{sec:overview}: For $\Delta=0$, the first optical mode contains the information on $\hat{X}_m$ in the phase quadrature, $\hat{p}_1=i(\hat{c}_1^\dagger-\hat{c}_1)/\sqrt{2}$, the quadrature that does \textit{not}  exert a radiation pressure force on the mechanical oscillator. Through the feedback loop, the $\hat{c}_1$ mode will be fed into the second, $\hat{c}_2$ mode.  In order for the feedback to be effective, the phase quadrature of the $\hat{c}_1$ mode has to be fed into the amplitude quadrature of the $\hat{c}_2$ mode, $\hat{x}_2$, the quadrature that does couple to the mechanical oscillator. For this to occur, the field has to be displaced by the auxiliary local oscillator, such that $\varphi=\pi/2$.

Furthermore, Eqs.~\eqref{eq:dampshift} show that by tuning the delay, the feedback can either result in a frequency shift or damping. Since for $\tau\gamma_m\ll1$ we have approximately
\begin{equation}
\label{eq:xttau}
\hat{X}_m(t-\tau) \simeq \cos(\Omega_{m}\tau)\hat{X}_m(t)-\sin(\Omega_{m}\tau)\hat{P}_m(t),
\end{equation}
we can see that the delay determines which quadrature, $\hat{X}_m$ or $\hat{P}_m$, is being fed back to the oscillator. In the limit of no delay, we are feeding back a force proportional to the position, resulting in a strong frequency shift. Maximal damping can be achieved by feeding back a force proportional to the momentum, which occurs at $\Omega_m\tau = \pi/2$. We note that the induced damping can become negative as $\gamma_m+\Gamma_m<0$, in which scenario the system becomes unstable and our linearized description fails.

{It is instructive to carry out a similar analysis in frequency space. This is done in App.~\ref{app:langfreq}.}

\begin{figure}[t]
    \centering
    \includegraphics[scale=1]{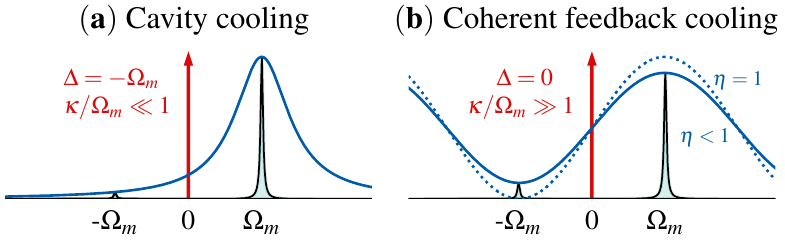}
    \caption{Sideband picture of optomechanical cooling. \textbf{(a)} $S_\mathrm{fb}(\omega)$ in the standard optomechanical cavity cooling for $\kappa/\Omega_m\ll1$ and $\Delta=-\Omega_m$. The Stokes processes are suppressed because they lie outside the cavity resonance. \textbf{(b)} $S_\mathrm{fb}(\omega)$ in the coherent feedback scenario, for $\kappa/\Omega_m\gg 1$ and $\Delta=0$, see Eq.~\eqref{eq:fbnoise}. The Stokes processes in the two optical passes interfere destructively, such that the pump beam predominantly extracts phonons and cools the mechanical oscillator. The dotted line corresponds to the ideal case in the absence of losses with $\eta=1$.}
    \label{fig:sidebands}
\end{figure}

{\subsection{Sideband picture}
Just as for cavity dynamical backaction cooling \cite{aspelmeyer2014}, we may develop a sideband picture and write the damping induced by the coherent feedback as
\begin{equation}
    \label{eq:dampstokes}
    \Gamma_m = A^--A^+,
\end{equation}
where $A^\pm= S_{\rm fb}(\mp\Omega_m)$ denote the rates for the Stokes $(A^+)$ and anti-Stokes $(A^-)$ processes which are determined by the spectral density of the feedback noise
\begin{equation}
    \label{eq:fbnoisesimple}
    S_{\rm fb}(\omega) = \frac{8}{\kappa}\left[\frac{g_1^2+g_2^2}{2}-\sqrt{\eta}g_1g_2\cos(\varphi+\omega\tau)\right],
\end{equation}
where the last equality holds for the unresolved sideband regime on cavity resonance [see Eq.~\eqref{eq:fbnoise} for the full expression]. Equation \eqref{eq:fbnoise} is plotted in Fig.~\ref{fig:sidebands}\,(b).
}

{The phonon number of the mechanical oscillator may then be written as
\begin{equation}
    \label{eq:number}
    \bar{n}_m = \frac{\gamma_m n_{\rm th}+A^+}{\gamma_m+\Gamma_m},
\end{equation}
where $n_\mathrm{th}=n_{B}(\Omega_m)$ denotes the thermal occupation. From the last equation we find that in order to reach a phonon number close to zero, two conditions need to be met. First, the quantum regime, where thermal excitations can be neglected, needs to be reached. This requires a large quantum cooperativity 
\begin{equation}
    \label{eq:quantumcoop}  
    \mathcal{C}_{\rm qu} = \frac{4g_1g_2}{\kappa\gamma_m n_\mathrm{th}}\gg 1.
\end{equation}
Note that the quantum cooperativity usually only includes a single optomechanical coupling. Here, the product of the two couplings $g_1$ and $g_2$ is relevant.}

The second condition to reach the ground state is a suppression of the Stokes processes (i.e., $A^+\ll A^-$), such that phonons are predominantly absorbed by Anti-Stokes scattering into the pump beam. For cavity dynamical backaction cooling in the sideband resolved regime, this is achieved by detuning the drive by an amount $\Delta=-\Omega_m$, see Fig.~\ref{fig:sidebands}\,(a). In contrast, in our coherent feedback scheme, which operates in the unresolved sideband regime, this suppression results from the interference between scattering processes in the first and second optical passes, which results in a frequency-dependent modulation of the noise spectral density $S_{\rm fb}(\omega)$ as illustrated in Fig.~\ref{fig:sidebands}\,(b) and thus a suppression of quantum backaction heating.

In the unresolved sideband limit and on resonance, we find the {minimum phonon number
\begin{equation}
    \label{eq:numberlim}
   \bar{n}_m = \frac{A^+}{\Gamma_m}=\frac{A^+}{A^--A^+}\geq \frac{1-\sqrt{\eta}}{2\sqrt{\eta}},
\end{equation}}
where the lower limit is reached for {a large quantum cooperativity}, $\Omega_m\tau=\varphi$, and $g_1=g_2$. Interestingly, this lower bound has the same form as the limit of measurement-based feedback cooling \cite{bowen2015,rossi2018}, with the efficiency of the feedback loop $\eta$ replacing the measurement efficiency.
As discussed in more detail in Sec.~\ref{sec:meascool}, {we can draw an analogy between coherent feedback cooling and measurement-based cooling: the coherent feedback cooling can, in the ideal limit, be understood as a measurement-free implementation where the readout signal is coherently transformed into a feedback actuation.  In App.~\ref{ap:coolingperformance} we compare the performance of coherent feedback cooling with standard cavity backaction cooling for different cavity regimes. }

\section{Coherent feedback control of a nanomechanical membrane}

Our experimental setup consists of a mechanical oscillator inside a cavity in a cryogenic environment provided by a low-noise liquid-Helium flow cryostat.  The mechanical oscillator is the (2,2) square drum mode of a silicon nitride membrane \cite{thompson2008} with a vibrational frequency $\Omega_m = 2\pi\times\SI{1.9}{MHz}$. The membrane is surrounded by a silicon phononic bandgap structure which shields this mode, leading to intrinsic quality factors that range from  $Q = \Omega_m/\gamma_m = \SI{1.9e6}{}$ at room temperature to $Q = \SI{3.2e6}{}$ at $\SI{20}{K}$. The membrane is placed inside a 
single-sided optical cavity of free spectral range \SI{150}{GHz}, finesse $\mathcal{F}= 1200$ and linewidth $\kappa = 2\pi\times\SI{55}{MHz}$, such that the optomechanical system operates in the unresolved sideband regime $\kappa\gg\Omega_m$. The bare optomechanical coupling strength is $g_0 =2\pi\times \SI{160}{Hz}$, calibrated via a phase modulation tone \cite{gorodetksy2010a}.

The overall efficiency of the feedback loop is determined by a combination of different losses that accumulate along the optical path. Following the optical path illustrated in Fig.~\ref{fig:SetupPhaseSketch}\,(a), for the first beam we have to consider the finite cavity incoupling efficiency $\eta_1$ = 0.91.  For the second pass, it includes the unavoidable loss at the beamsplitter that combines the auxiliary local oscillator $\hat{a}_{\mathrm{aux}}$ and the back reflection of the first beam $\hat{a}^{\mathrm{out}}_{1}$, which has a splitting ratio $\eta_{\mathrm{aux}}$ = 0.87 {in our experiment}. {Note that this loss can be made arbitrarily small in principle by using a strongly unbalanced beam splitter and higher incoming optical power for $\hat{a}_{\mathrm{aux}}$.}
Additionally, there is a cumulative loss due to the propagation in the optical fiber and other optical elements{, leading to a transmission efficiency $\eta_T \simeq 0.3 $} together with the cavity incoupling efficiency of the second beam in orthogonal polarization {$\eta_2$ = 0.9}. As discussed in App.~\ref{ap:langevin}, these losses can be fully taken into account by the overall efficiency of the feedback loop $\eta=0.22$ and by an appropriate rescaling of the average displacements. { Since the powers are measured in front of the cavity, $P_1$ is measured directly but the measured auxiliary power is given by $\tilde{P}_\mathrm{aux} = (1-\eta_\mathrm{aux})\eta_T{P}_\mathrm{aux}$}. 
In the following experiments, we use the delay and the phase of the coherent feedback loop as the tuning knobs that allow us to control the mechanical state of the membrane,  as described by Eq.~\eqref{eq:dampshift}. This  shows up in the recorded mechanical power spectral densities as a change of both the mechanical linewidth and the oscillation frequency, which we extract from Lorentzian fits to our data. 

\subsection{Control via the loop delay}
In a first set of experiments, we study the effect of delay alone without an auxiliary local oscillator in the feedback loop. We generate different delays between the two interactions with the mechanical oscillator by sending the light after the first pass through optical fibers of different lengths. Interesting situations arise once the delay is significant, i.e.\ of order $\Omega_m \tau \sim$ $\pi $. We investigate the effect generated by different delays starting from a fiber length of \SI{2}{m} going up to \SI{80}{m}, which corresponds to $\Omega_m \tau = 0.07 \pi $ up to $\Omega_m \tau =  1.55 \pi $. 

At zero detuning, the motion of the membrane is imprinted purely as a phase modulation onto the output light such that in the absence of the auxiliary local oscillator this results in $\varphi =  \pi$ (due to the back reflection from the cavity) and we expect no effect from the coherent feedback loop  [cf.~Eqs.~\eqref{eq:dampshift}]. With a finite detuning however, a phase shift $\varphi \neq \pi$ is introduced even without any auxiliary local oscillator  [cf.~Eqs.~\eqref{eq:means}]. Therefore, in that case, the motion is imprinted onto both the amplitude and phase quadratures and the effect of different delays due to feedback becomes apparent. Additionally, the standard cavity dynamical backaction effects that are not captured by Eqs.~\eqref{eq:dampshift} modify the frequency shift and damping rate, see App.~\ref{ap:theo} for details. 

\begin{figure}[h]
    \centering
    \includegraphics[scale=1]{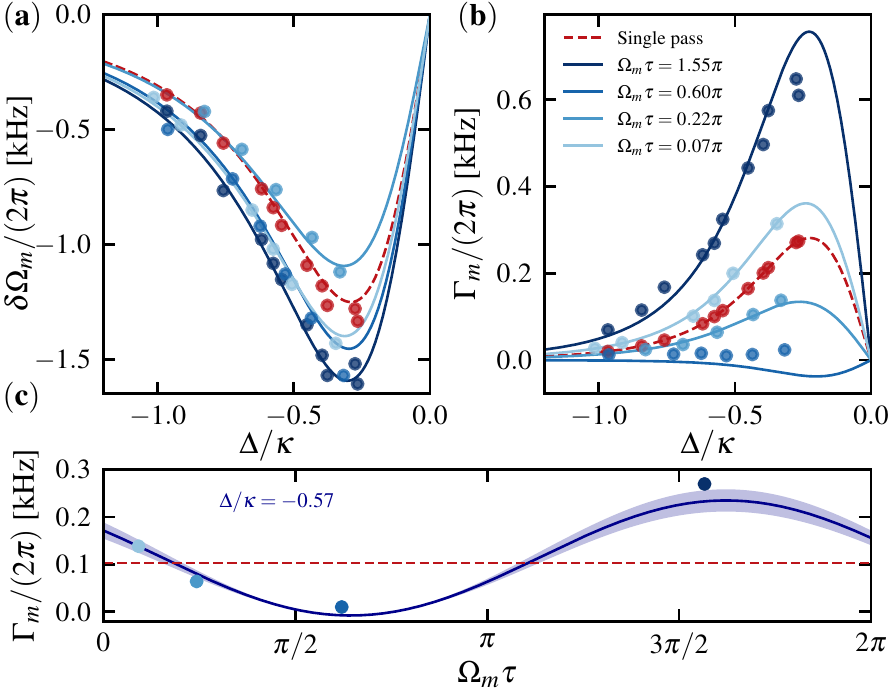}
    \caption{Mechanical frequency shift \textbf{(a)} and damping rate \textbf{(b)} as a function of the cavity detuning for different feedback delays{, with $\hat{a}_\mathrm{aux} = 0$}. The data points correspond to the results of Lorentzian fits to the mechanical power spectral density. The solid lines correspond to the theoretical predictions in Eqs.~\eqref{eq:delomom} and \eqref{eq:gammaom} evaluated at $\omega=\Omega_m$ with no free parameters. The detuning axis is calibrated from the measured linewidth in the single-pass interaction (dashed red line). \textbf{(c)} The mechanical linewidth at a detuning of $\Delta / \kappa$ = -0.57 for the different fibres and respective delays.}
    \label{fig:CoherentFeedbackDelay}
\end{figure}

Figure~\ref{fig:CoherentFeedbackDelay} shows experimental data where we study the effect of different feedback delays while scanning the detuning for an input power $P_1=\SI{60}{\micro W}$. The coherent feedback onto the mechanical oscillator results in both in a shift of the mechanical frequency [Fig.~\ref{fig:CoherentFeedbackDelay}(a)] and in a broadening or narrowing of the mechanical linewidth [Fig.~\ref{fig:CoherentFeedbackDelay}(b)], leading to damping or driving, respectively. This is consistent with a picture in which the membrane motion couples via the light to a delayed version of itself, leading to feedback forces $F_\mathrm{fb}\propto \pm P_m(t)$ for certain delays as shown in Eq.~\eqref{eq:dampharmx} and in Fig.~\ref{fig:CoherentFeedbackDelay}(c). 

Indeed, we observe that for a delay close to $\Omega_m\tau\sim\pi/2$ (i.e. a quarter of the oscillation period) the coupling is mostly proportional to $+P_m$ and we observe driving (narrowing of the linewidth) even with a red detuned beam. Half a period later, for $ \Omega_m\tau \sim 3\pi/2$,  the feedback force is mostly proportional to $ -P_m$ and the  motional damping is amplified by more than a factor 3 as compared to a single interaction, leading to additional cooling of the mechanical oscillator. 
Finally, we see that for the smaller delays $\Omega_m\tau\sim 0$, the effect of the second interaction on the broadening  is small $\Gamma_m\simeq 0$, since the feedback force, in this case, is mostly $\propto X_m$. 

The agreement with the theoretical predictions in Eqs.~\eqref{eq:delomom} and  \eqref{eq:gammaom} (solid lines in Fig.~\ref{fig:CoherentFeedbackDelay}) is excellent. The theory lines for the feedback interaction contain no free parameters. The detuning axis is calibrated from the recorded linewidths in the single-pass interaction under the effect of the standard cavity dynamical backaction \cite{aspelmeyer2014} and can be extracted with an uncertainty of $\pm 5\%$.

\subsection{Control via the loop phase}

{Control over the feedback phase is a handy knob in a feedback platform, allowing to modify the effect of the feedback on the system under control}. Here, we investigate how the loop phase modifies the membrane motion at a fixed delay and cavity detuning. 
As previously discussed, this phase allows us to control the amount of motional information that is transferred onto the amplitude quadrature of the second interaction beam, thereby maximizing or minimizing the feedback force on the membrane, as well as the overall sign of the interaction. Experimentally, we {vary} the loop phase $\varphi$ by adjusting the phase of the auxiliary local oscillator {$\phi=\arg(\alpha_\mathrm{aux}/\alpha^\mathrm{out}_1)$}, which is selected and stabilized by locking at a specific angle of the interferometric signal between a small leak of $\alpha^\mathrm{out}_1$ and $\alpha_\mathrm{aux}$ [see Fig.~\ref{fig:SetupPhaseSketch}].  

\begin{figure}[ht]
    \centering
    \includegraphics[scale=1]{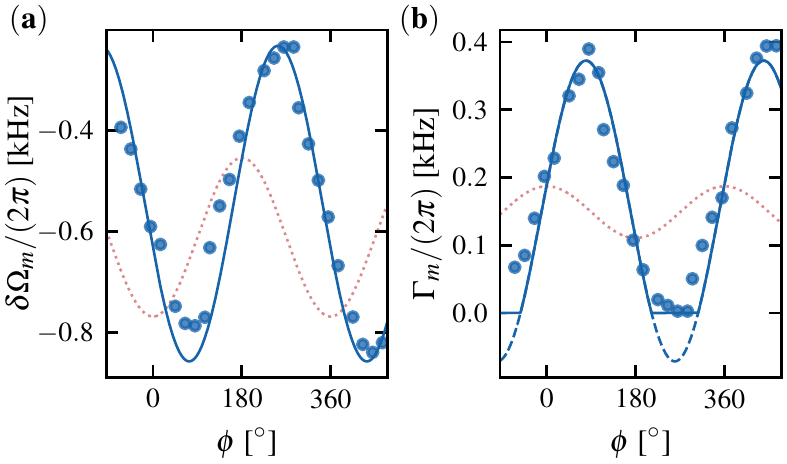}
    \caption{Mechanical frequency shift \textbf{(a)} and damping rate \textbf{(b)} as a function of the phase of the auxiliary local oscillator. The red dotted lines correspond to the broadening expected in the absence of feedback, but with an equivalent power in a single beam. The solid lines correspond to Eqs.~\eqref{eq:delomom} and \eqref{eq:gammaom} evaluated at $\omega=\Omega_m$ with no free parameters. 
    The  solid line in \textbf{(b)} is set to zero for negative values (dashed line), where the mechanical oscillator is driven by the feedback.}
    \label{fig:CoherentFeedbackPhase}
\end{figure}

The measured frequency shifts and linewidths are shown in Fig.~\ref{fig:CoherentFeedbackPhase}. In this measurement, the delay is held constant at $ \Omega_m\tau\sim 0.07 \pi $ and the detuning at $\Delta / \kappa = -0.2 $, the input powers were set to $\tilde{P}_{1}  = \SI{20}{\micro W} $ and $\tilde{P}_{\mathrm{aux}} = \SI{3}{\micro W}$. This detuning is experimentally chosen such that the amount of standard cavity dynamical backaction cooling is strong. This allows us to show the coherent feedback effect by scanning the full range $2 \pi $ of the loop phase without running into instabilities when approaching negative effective linewidths. In practice, when we reach this unstable regime [blue dashed lines in Fig.~\ref{fig:CoherentFeedbackPhase}\,(b)], the system is driven and the measured linewidth is close to zero.

Scanning the phase, we observe that both the resonance frequency and the linewidth can be modified to either higher or lower values compared to the optical spring and broadening that occur even without the coherent feedback. We exploit this aspect in the next section to optimally feedback cool the mechanical oscillator. Here again, we find an excellent agreement between the experimental data points and the theoretical prediction in Eqs.~\eqref{eq:delomom} and \eqref{eq:gammaom} with no free parameters. 

\section{Coherent feedback cooling below the dynamical backaction limit}

\begin{figure*}[t]
    \centering
    \includegraphics[width=\textwidth]{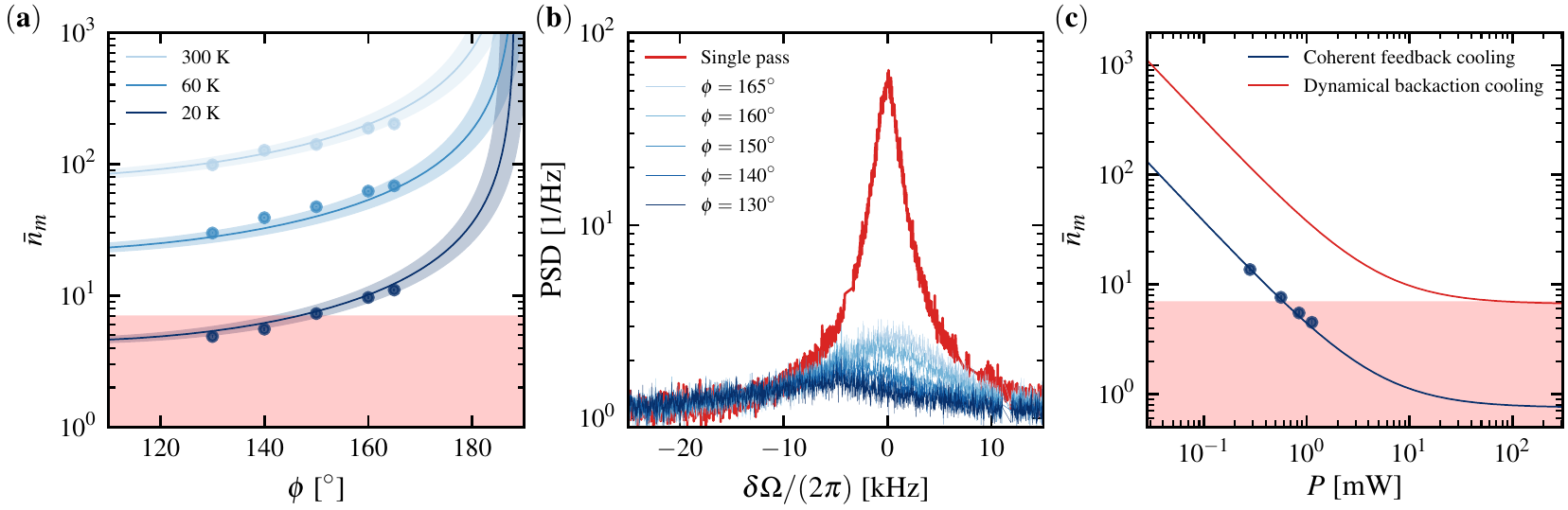}
    \caption{Coherent feedback cooling below the dynamical backaction limit. \textbf{(a):} Phonon occupation plotted as a function of the phase of the auxiliary local oscillator at different cryostat temperatures for input powers ${P}_{1} = \SI{0.4}{mW}$ and $\tilde{P}_\mathrm{aux} = \SI{1.2}{mW}$ and detuning $\Delta/\kappa = -0.35$.  The red shaded area indicates the limit of cavity dynamical backaction cooling.  The shaded areas around the theory lines correspond to a $\pm5\%$ uncertainty in the inferred detuning. The error bars take into account both the numerical uncertainty from the fit of the raw data and the propagation of uncertainties from the calibrated quantities and are small compared to the point size.  \textbf{(b):} Shot-noise normalized mechanical power spectral densities corresponding to the data points at $T=\SI{20}{K}$ in \textbf{(a)}, the frequency origin is centered so that the $\delta\Omega=0$ corresponds to the single-pass mechanical resonance frequency.
    \textbf{(c):} Coherent feedback cooling compared to standard cavity dynamical backaction cooling.  Blue line and data points: Phonon occupation at a constant phase $\phi = 130^{\circ}$, scanning the total input power resulting from the double-pass interaction $P = {P}_1 (1+\eta_T\eta_\mathrm{aux})  +\tilde{P}_{\mathrm{aux}} + 2 ({\eta_\mathrm{aux} \eta_{2} {P}_1 \tilde{P}_{\mathrm{aux}} })^{1/2} \cos{(\phi)} $ while keeping the ratio $\tilde{P}_\mathrm{aux}/{P}_1 = 3$ fixed, at a detuning $\Delta/\kappa = - 0.35 $. 
    Red line: cooling by standard cavity dynamical backaction given the same total input power $P$ at the optimal detuning $\Delta/\kappa = - 0.5 $ for dynamical backaction.}
    \label{fig:CoherentFeedbackCooling}
\end{figure*}

For cavity optomechanical systems within the so-called resolved sideband regime, it has been established theoretically and demonstrated in multiple platforms that a red-detuned drive allows cooling the mechanical oscillator close to its motional ground state \cite{aspelmeyer2014}. Outside this regime, cavity dynamical backaction cooling to the ground state is no longer attainable and the most widely used  cooling technique is measurement-based feedback \cite{mancini1998,genes2008}, where the optical signal is measured and converted into a classical electronic signal that drives the feedback actuator. 

Here, we exploit the control provided by the coherent feedback loop demonstrated in the previous section to cool the membrane close to the ground state in the unresolved sideband regime. The available tuning knobs are the loop phase $\varphi$ and delay $\tau$ as well as the detuning $\Delta$ and the powers of the first and auxiliary local oscillators $P_1$ and $P_\mathrm{aux}$. 
In standard cavity cooling, the minimal number of phonons achievable in the unresolved sideband regime is bounded by $\kappa/(4\Omega_m)$ (cf.~App.~\ref{ap:theo_unr}), which in our case corresponds to about 7 phonons. In order to reach this dynamical backaction limit with our mechanical quality factor, we would need a laser power on the order of $\SI{100}{mW}$ [see Fig.~\ref{fig:CoherentFeedbackCooling} (c)]. The coherent feedback loop dramatically relaxes this power constraint and we are able to cool the motion of the membrane below the dynamical backaction limit, approaching the ground state. 

In Fig.~\ref{fig:CoherentFeedbackCooling}\,(a) we show experiments where we reach our lowest membrane phonon occupation by scanning the experimental feedback loop phase. We present measurements at different cryostat temperatures, where the delay is set to $\Omega_m\tau\sim \pi/4$ and the detuning is kept fixed at $\Delta/\kappa=-0.35$. The powers for the first and auxiliary local oscillators are set to $\SI{400}{\micro W}$ and $\SI{1.2}{mW}$, respectively.
With these experimental parameters, the feedback loop drives the mechanical oscillator towards a state with phonon occupation of $\bar{n}_m = 4.89\pm 0.14$  phonons  for a cryostat temperature of \SI{20}{K}, reaching a phonon number below the theoretical limit of cavity dynamical backaction cooling for our membrane-cavity assembly. The coherent feedback cooling rate is $\Gamma_{m}>10 \, \Gamma_\mathrm{dyn}$, where $\Gamma_\mathrm{dyn}$ is the cooling rate of dynamical backaction cooling at the same power. 

In these experiments, the membrane phonon occupation is determined from the area of the mechanical power spectral density, recorded via phase-sensitive homodyne detection. By determining the reduction in area with respect to a single interaction  [see Fig.~\ref{fig:CoherentFeedbackCooling}\,(b) and App.~\ref{ap:calibration}] we extract $\bar{n}_m$.

We note that higher powers, smaller detunings and slightly smaller phases should further reduce the final number of phonons, but these parameter regimes were not accessible to us due to technical instabilities related to the cavity lock. Similarly, the optimal delay $\Omega_m\tau\sim \pi/2$ could not be implemented, most likely due to { the increased} phase noise in longer fibers. {We also note that, due to the cavity detuning we use, incident phase noise will be rotated into the light amplitude quadrature, which can in turn heat the  mechanical oscillator mode. }
{Furthermore, the membrane sustains a multitude of higher-order mechanical modes which experience coherent feedback with different phase shifts depending on their resonance frequencies. For those modes with sufficiently strong optomechanical coupling \cite{nielsen2017}, the feedback effect might turn into amplification instead of cooling and drive the cavity to an unstable regime. Although this could be a potential issue, we did not observe any effect in our measurements. }

\section{Comparison with measurement-based feedback cooling} 
\label{sec:meascool}
It is illustrative to compare our coherent feedback scheme to well-known measurement-based feedback (mf) schemes for the specific task of cooling mechanical motion~\cite{mancini1998,genes2008,rossi2018, tebbenjohanns2021, delic2020}. In measurement-based feedback, the all-optical loop is replaced by an {optoelectronic} loop. An estimate of the mechanical displacement is obtained from a measurement of the phase quadrature of the output light from cavity mode $\hat{c}_1$, and then a mechanical force proportional to the derivative of the estimated displacement is applied. In the case where the feedback force is optomechanically actuated, the electronic signal modulates the amplitude of a laser driving the second cavity mode $\hat{c}_2$, which results in a time-dependent coupling $g_2$. Irrespective of the particular implementation of the feedback force, the effect of the feedback can be effectively modeled by replacing the coupling to the second cavity in the equation of motion for $\hat P_{m}$ with a feedback force, entering Eq.~\eqref{eq:langevinmech} as a convolution term $F_\mathrm{mf}(t)=(h_\mathrm{mf}* p_{1}^\mathrm{est})(t)$, where $h_\mathrm{mf}$ denotes the feedback transfer function and 
$p_{1}^\mathrm{est}=p^\mathrm{out,\,\eta_\mathrm{det}}_\mathrm{1}/\sqrt{\eta_\mathrm{det}\kappa}$. Notice that the measurement is limited by the finite quantum efficiency of the detector $\eta_\mathrm{det}$, i.e., $p^{\mathrm{out}, \,\eta_\mathrm{det}}_1=\sqrt{\eta_\mathrm{det}}p^\mathrm{out}_1-\sqrt{1-\eta_\mathrm{det}}p_0$, where $p_0$ is an uncorrelated vacuum field.
For concreteness, here we focus on the case of the so-called cold damping scheme~\cite{genes2008}, while a more general treatment is provided in App.~\ref{ap:measfb}. Cold damping is characterized by the following spectral filter function
\begin{equation}\label{eq:Hfb}
h_\mathrm{mf}(\omega)=-i\frac{g_\mathrm{mf}\,\omega }{1-i{\omega}/{\omega_\mathrm{mf}}}\,,
\end{equation}  
which is expressed in terms of the bandwidth $\omega_\mathrm{mf}$ and by the dimensionless quantity $g_\mathrm{mf}$, which quantifies the feedback gain. The bandwidth $\omega_\mathrm{mf}$ describes the finite time response of the feedback, while any explicit delay in the feedback loop is neglected~\cite{sommer2020}. From Eq.~\eqref{eq:Hfb} we can already understand the regime of interest for feedback by observing that $\arg(h_\mathrm{mf})=-\arctan({\omega_\mathrm{mf}}/{\omega})$. For $\omega_\mathrm{mf}\gg\omega\approx \Omega_{m}$, the argument tends to $-\pi/2$, so that the feedback force becomes proportional to momentum, thus providing damping of the mechanical motion. Therefore, the relevant regime for feedback cooling is that of large feedback bandwidth. 

Similar to coherent feedback cooling, the effect of measurement-based feedback cooling is fully taken into account by introducing a modified mechanical frequency and a modified damping rate, which are given by
\begin{align}
\Gamma_\mathrm{mf}&= \frac{2\Omega_{m} g_1 g_\mathrm{mf} \omega_\mathrm{mf}[({\kappa}/{2})\,\omega_\mathrm{mf}-\Omega_{m}^2]}{\left[ \left( {\kappa}/{2}\right)^2+\Omega_{m}^2\right](\Omega_{m}^2+\omega_\mathrm{mf}^2)} \,,\label{eq:GammaFb} \\
\delta\Omega_\mathrm{mf}&= \frac{\Omega_{m}^2 g_1 g_\mathrm{mf} \omega_\mathrm{mf}\,({\kappa}/{2}+\omega_\mathrm{mf})}{\left[ \left( {\kappa}/{2}\right)^2+\Omega_{m}^2\right](\Omega_{m}^2+\omega_\mathrm{mf}^2)} \,. \label{eq:OmegaFb} 
\end{align} 
In the relevant limit $\omega_\mathrm{mf},\,\kappa\gg \Omega_{m}$,
the residual phonon occupation, as obtained from the corresponding noise {power} spectral density, is given by
\begin{align}
\label{eq:Nfeedback}
\bar{n}_{m,\mathrm{mf}} &=\frac{g_1}{ g_\mathrm{mf}\Omega_{m}} +\frac{g_\mathrm{mf}\Omega_{m}}{16 g_1 \eta_\mathrm{det}} -\frac12 
\ge \frac{1-\sqrt{\eta_\mathrm{det}}} {2\sqrt{\eta_\mathrm{det}}}\,,
\end{align}
where the inequality is saturated for a value of the feedback gain $g_\mathrm{mf}= 4 \sqrt{\eta_\mathrm{det}}g_1/ \Omega_{m}$. 
Remarkably, the above expression has the same form as the residual occupation of coherent feedback cooling, given by Eq.~\eqref{eq:numberlim}. 

To better appreciate this correspondence, we can evaluate Eqs.~\eqref{eq:dampshift} and \eqref{eq:numberlim} for $\Omega_{m}\tau=\varphi=\pi/2$, i.e.\ the parameters that result in optimal cooling. We find
\begin{equation}
\begin{aligned}
&\Gamma_m = 16\sqrt{\eta}\frac{g_1g_2}{\kappa},\hspace{.5cm}\delta\Omega_m = 0,\\&
\bar{n}_m =\frac{g_1}{4\sqrt{\eta} g_2} +\frac{g_2}{4\sqrt{\eta} g_1} -\frac12\,.
\end{aligned}
\end{equation}
In the limit $\omega_\mathrm{mf},\,\kappa\gg \Omega_{m}$, we can recover these expressions from Eqs.~\eqref{eq:GammaFb}, \eqref{eq:OmegaFb}, and \eqref{eq:Nfeedback} upon setting $g_{\rm mf} = 4\sqrt{\eta}g_2\Omega_m$ and $\eta_{\rm det}=\eta$. We, therefore, conclude that in this limit the two schemes are equivalent. In App.~\ref{ap:measfb} we show that this equivalence can be extended to arbitrary delays $\tau$ and linewidths $\kappa$. 

We note that beyond cooling, {in} applications where both light quadratures might play a role, coherent and measurement-based feedback control are not equivalent anymore \cite{zhang2017}. 

\section{Conclusions \& Outlook}
We implemented an all-optical coherent feedback platform to control the motion of a mechanical oscillator and demonstrated full control via the parameters of the feedback loop, namely the phase and the delay. 
We showed theoretically that this scheme can be used for ground-state cooling in the unresolved sideband regime  without the need for measurements. We demonstrated experimentally that even with a moderate mechanical $Q-$factor, we can beat the theoretical lowest-phonon number limit of cavity dynamical backaction cooling in the unresolved sideband regime, {with only 1$\%$ of the optical power required for the latter}.  
In contrast to previous proposals, where feedback is performed on the optical cavity mode \cite{harwood2021, guo2022}, we perform feedback directly on a mechanical oscillator mode {, using orthogonal polarizations for first and second passes. But our scheme does not rely on the availability of same-frequency orthogonal cavity modes: When experimentally possible, one could use another longitudinal mode of the cavity.  Alternatively, the scheme could also be implemented via a loop that is opened or closed by an optical switch,  with a switching rate $1/(2\tau)$, in such a way that first and second passes are never at the same time in the cavity.} As such, the double-pass scheme can be adapted to a variety of different physical systems and is not restricted to optomechanics.
{In the present configuration, ground state cooling would be achievable by improving the thermal conductivity of the membrane support to ensure thermalization at $\SI{4.2}{\kelvin}$. However, the most straightforward improvement would consist in using a mechanical resonator with a higher quality factor \cite{tsaturyan2017}.}

The beauty of coherent feedback lies in its potential for processing non-commuting observables \cite{hamerly2012}. In the unresolved sideband regime, coherent feedback opens up the possibility to generate self-interactions and mechanical squeezing \cite{karg2019}, photon-phonon entanglement \cite{guo2022}, or to enhance optical nonlinearities at the single-photon level \cite{wang2017a}. 
Our scheme could also be exploited in the sideband resolved regime, where optomechanical couplings of the form $\hat{c}_j\hat{B}+\hat{B}^\dagger\hat{c}_j^\dagger$ can be designed, with $\hat{B}$ being a raising or lowering operator of the mechanical oscillator. Such couplings are sensitive to both quadratures and potentially allow for creating non-classical mechanical states using coherent feedback \cite{wiseman1994}.

In contrast to measurement-based control, coherent feedback avoids the incoherent addition of feedback and measurement noise, making it a key technique in the low phonon-number regime \cite{hamerly2012}. In particular, {the interference of optical input noise can be tuned} to realize backaction cancellation \cite{karg2019, tsang2010}. This makes coherent feedback a promising candidate for sensing applications, where such backaction cancellation is highly desirable.

Coherent feedback thus opens up new approaches for engineering the dynamics of quantum systems with potential applications for quantum technology, measurement and control as well as quantum thermodynamics.

\begin{acknowledgements}
We thank Chun Tat Ngai for discussions. This work was supported by the project “Modular mechanical-atomic quantum systems” (MODULAR) of the European Research Council (ERC) and by the Swiss Nanoscience Institute (SNI). M.B.A. acknowledges funding from the European
Union’s Horizon 2020 research and innovation programme under the Marie Sk\l odowska-Curie
grant agreement N°101023088. P.P.P. and M.B. acknowledge funding from the Swiss National Science Foundation (Eccellenza Professorial Fellowship PCEFP2\_194268).
\end{acknowledgements}

\appendix

\section{Phonon number calibrations}
\label{ap:calibration}
\subsection{Phonon number via homodyne detection}

In this section, we detail the calibration of the phonon number, which is determined by performing a homodyne measurement on a leak of the first beam interacting with the membrane [see Fig.~\ref{fig:SetupPhaseSketch}(a)]. This is done at room temperature, where we know the membrane is thermalized to the environment.  

In the frequency domain, we can write the field $\hat{c}_1$ inside the cavity as \cite{bowen2015}
    \begin{equation}
        \hat{c}_1(\omega) = -\chi_c(\omega)\left[\sqrt{\kappa\eta_1}\hat{a}^\mathrm{in}_1(\omega) + i\sqrt{2}g_1\hat{X}_{m}(\omega)\right],
    \end{equation}
where $\eta_1$ is the cavity incoupling efficiency and $\chi_c(\omega)^{-1} = \kappa/2 - i(\Delta +\omega ) $ the cavity susceptibility. 

The output phase quadrature of the light $\hat{P}_L$ is related to the one inside the cavity $\hat{p}_1$ via the input-output relation $\hat{P}_L=\hat{P}_1^\mathrm{in}+\sqrt{\kappa\eta_1}\hat{p}_1$. 

In the thermal-noise-dominated regime, we can neglect the  term $\hat{P}_1^\mathrm{in}$. Setting $g_1=g_0\,\chi_c(0)\sqrt{\kappa\eta_1}\alpha^\mathrm{in}_1$ we can express $\hat{P}_L$ in terms of experimentally measurable quantities:
    \begin{equation}
       \hat{P}_L (\omega) =-\eta_1 g_0 \alpha^\mathrm{in}_1\,\mathcal{R}(\omega)\hat{X}_m(\omega),
    \end{equation}
with the cavity transduction factor
    \begin{equation}
     \mathcal{R} (\omega) = \kappa \,\big[ \chi_c(0) \chi_c(\omega) + \chi^{\ast}_c(0) \chi^{\ast}_c(-\omega)\big] 
     \overset{\Delta = 0 }{=} \frac{8}{\kappa}.
    \end{equation}

The thermal occupation number $\bar{n}_m$ is then calibrated by measuring this light quadrature. For this, we beat the output beam with a strong local oscillator with power $P_\mathrm{LO} = \hbar\omega_L \lvert \alpha_\mathrm{LO} \rvert^2 \gg P_1$ and use a Mach-Zehnder interferometer to perform homodyne detection. By means of a piezoelectric mirror in the local oscillator arm, we can scan the homodyne angle $\theta$, and the recorded output voltage is given by:
    \begin{equation} 
     \hat{D}(\omega) = \sqrt{2} \alpha_{\mathrm{LO}}  \big[ \cos{(\theta)} \hat{X}_L (\omega)  + \sin{(\theta)} \hat{P}_L ( \omega)  \big].
    \end{equation}
The DC signal of the interference is given by $D(\omega=0) = 2 \alpha_{\mathrm{LO}} \alpha^\mathrm{in}_1\cos\theta$, whose amplitude $D_0=2 \alpha_{\mathrm{LO}} \alpha^\mathrm{in}_1$ can now be used to calibrate the membrane signal. Locking the interferometer at $\theta=\pi/2$ we are sensitive to the phase quadrature $\hat{P}_L$ encoding the membrane signal $\hat{X}_m$
    \begin{equation}
          \hat{D}_{\theta = \frac{\pi}{2} } (\omega) = \sqrt{2} \alpha_{\mathrm{LO}} \alpha^\mathrm{in}_1  \eta_1 g_0 \mathcal{R} (\omega)  \hat{X}_m (\omega).
    \end{equation}

From this signal $\hat{D}_{\theta = \frac{\pi}{2} } (\omega) $, we can compute the detected power spectral density (PSD) $S_{DD}(\omega)$
    \begin{equation}
    \label{eq:SDD}
        S_{DD}(\omega)=\frac{1}{2}D_0^2\left[\eta_1g_0 \,|\mathcal{R}(\omega)|\right]^2S_{XX}(\omega),
    \end{equation}
with the membrane displacement power spectral density  $S_{XX} (\omega)$.
On the other hand, the average number of phonons is related to the variance of the membrane displacement as
    \begin{align}
    \label{eq:PSD}
        \bar{n}_m + \frac{1}{2} = \langle \hat{X}^2_m(t)\rangle = 2 \int^{\infty}_{0} \bar{S}_{XX} (\omega)\, \frac{d \omega}{2 \pi},
    \end{align}
with the symmetrized PSD $\bar{S}_{XX} (\omega)$.
Therefore, combining Eqs.~\eqref{eq:SDD} and \eqref{eq:PSD} the number of phonons can be directly obtained from the recorded power spectral density as
    \begin{equation}
        		\bar{n}_{m} = \frac{4}{D_0^2\left[\eta g_0 \,|\mathcal{R}(\omega)|\right]^2}	\int_{0}^\infty \bar{S}_{DD}(\omega)\,\frac{d \omega}{2\pi}-\frac{1}{2}.
    \end{equation}

\subsection{Phonon number via area ratios}

An alternative way to determine the thermal occupation number of the mechanical oscillator consists in comparing the measured power spectral density area in the presence of feedback with the area obtained if only a single interaction takes place.

In a single-pass interaction, the phonon occupation can be estimated from the procedure outlined in the previous section or from the measured linewidth $\Gamma_m$ and knowledge of the environment temperature by $\bar{n}_\mathrm{calib}=n_\mathrm{th}(T)\gamma_m/(\gamma_m+\Gamma_m)$. We can now associate the area of the measured spectrum $A_{\mathrm{calib}}$ to an occupation $\bar{n}_\mathrm{calib}$. 

For this, we first obtain the membrane occupation in the presence of moderate cooling due to the optical field following standard optomechanical cooling theory measured as a change in linewidth to its voltage transduction in our PSD measurement  $ \bar{S}^{\mathrm{calib}}_{DD} (\omega)$. 
We can then use the ratio between the corresponding calibration area  $A_{\mathrm{calib}}$ and the computed occupation number $\bar{n}_{\mathrm{calib}}$  to determine the occupation number $\bar{n}_m$ due to our coherent feedback loop.

\begin{figure}[ht]
    \centering
    \includegraphics[scale=1]{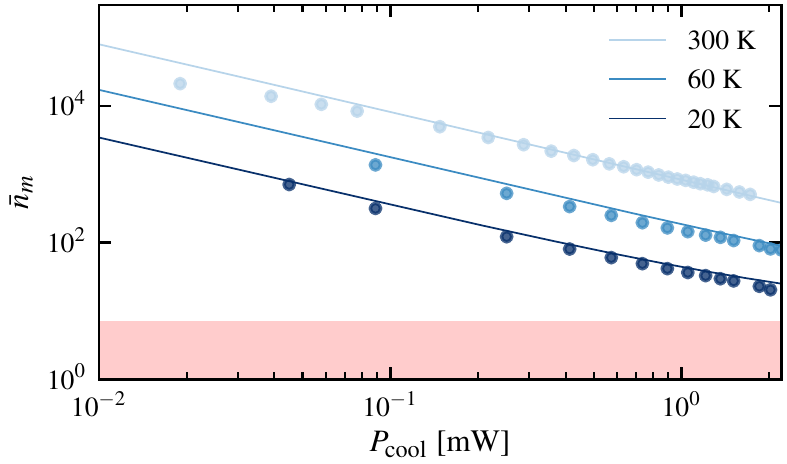}
    \caption{Standard dynamical backaction cooling. For different cryostat temperatures, the plot shows the phonon occupation as a function of cooling beam power.  The shaded area marks the cavity dynamical backaction limit. }
    \label{fig:CavityCooling}
\end{figure}

In the presence of additional cooling due to the feedback loop, the phonon number is then given by the ratio of areas

\begin{align}
    \bar{n}_m = \frac{\bar{n}_{\mathrm{calib}}}{A_{\mathrm{calib}} }  A_{DD}, 
\end{align}
with $A_{DD} $ the area of the measured PSD. 
For this to be accurate, it is essential to know the temperature $T$ at which the membrane is thermalized. To determine the actual thermalization temperature at different cryostat temperatures we use  standard optomechanical cooling experiments as shown in Fig.~\ref{fig:CavityCooling}. For each cryostat temperature, we increase the power of a red-detuned beam and measure the displacement PSD with a resonant probe beam in a homodyne detection scheme.  The ratio of the areas at different temperatures can be used to infer the temperature of the environment. If we compare the reduction in areas due to the optomechanical cooling to the theoretical expectation at those temperatures we find an excellent agreement.
Furthermore, this agreement allows us to exclude excess backaction through classical laser noise for the powers that we are employing. We find a satisfactory agreement between the two calibration procedures described in this and the previous section.  

\section{Theoretical description}
\label{ap:theo}
\subsection{Langevin equations including losses}
\label{ap:langevin}
In this section, we derive the Langevin equations provided in Eqs.~\eqref{eq:langevinmech}, \eqref{eq:langevinc1}, and \eqref{eq:langevinc2} from the non-linear optomechanical equations of motion including additional losses. We start by considering the Langevin equations for the mechanical oscillator
\begin{equation}
\label{eq:langevinmechap}
\begin{aligned}
\partial_t\hat{X}'_m(t) =& \Omega_m\hat{P}_m(t),\\\partial_t\hat{P}_m(t) =& -\Omega_m\hat{X}'_m(t)-\gamma_m\hat{P}_m(t)-\sqrt{2}g_0\sum_{j=1}^2\hat{C}_j^\dagger(t)\hat{C}_j(t)\\&-\sqrt{2}\,\hat{\xi}_{\rm th}(t).\\
\end{aligned}
\end{equation}
Here $\hat{C}_j$ denotes the annihilation operator in the cavity mode $j$ {and the radiation pressure force originates from the nonlinear optomechanical Hamiltonian $\hat H_{\mathrm{om},j}=\sqrt{2}\hbar g_0 \hat C_j^\dagger \hat C_j \hat X'_m$}~\cite{aspelmeyer2014}. We now write 
\begin{equation}
    \label{eq:displap}
    \hat{C}_j = \,e^{i\phi_j}\left(|\alpha_j|+\hat{c}_j\right),
\end{equation}
where $\alpha_j=\langle \hat{C}_j\rangle$. In the optomechanical coupling, we drop the non-linear term that is independent of $\alpha_j$, as it is negligible for large average displacements. This results in the equation for the momentum
\begin{equation}
\label{eq:langevinmechap2}
\begin{aligned}
\partial_t\hat{P}_m(t) =& -\Omega_m\left[\hat{X}'_m(t)+\delta X\right]-\gamma_m\hat{P}_m(t)\\&-2\sum_{j=1}^2g_j\hat{x}_j(t)-\sqrt{2}\hat{\xi}_{\rm th}(t),\\\delta X =& \sqrt{2}\frac{g_0}{\Omega_m}(|\alpha_1|^2+|\alpha_2|^2),
\end{aligned}
\end{equation}
where $\hat{x}_j=(\hat{c}_j^\dagger+\hat{c}_j)/\sqrt{2}$ as in the main text. To recover Eqs.~\eqref{eq:langevinmech} in the main text, we identify $\hat{X}_m=\hat{X}'_m+\delta X$, which also obeys $\partial_t\hat{X}_m = \Omega_m\hat{P}_m$. We note that $\delta X$ is of order one for the parameters in the experiment, implying that the mechanical oscillator exhibits an average displacement of the order of the zero-point fluctuations.

The Langevin equation for the cavity modes read
\begin{equation}
    \label{eq:langevinc1ap}
    \begin{aligned}
    \partial_t \hat{C}_j(t) = & \left(i\Delta-\frac{\kappa}{2}\right)\hat{C}_j(t)-i\sqrt{2}g_0\hat{C}_j\hat{X}'_m(t)\\&-\sqrt{\kappa}\hat{A}_{j}^{\rm in}(t),
    \end{aligned}
\end{equation}
where $\hat{A}_j^{\rm in} =\alpha_{j}^{\rm in}+\,e^{i\phi_j}\hat{a}_j^{\rm in}$ denotes the input field for the respective mode, with $\langle\hat{A}_j^{\rm in}\rangle =\alpha_{j}^{\rm in}$. Using Eq.~\eqref{eq:displap} and dropping the non-linear term in the optomechanical coupling, we find equations of motion for both the average displacements $\alpha_j$ and the operators describing fluctuations around those averages $\hat{c}_j$. For the averages, we find
\begin{equation}
    \label{eq:langevinc1apav}
    \partial_t \alpha_j(t) = \left(i\Delta+i\sqrt{2}g_0\delta X-\frac{\kappa}{2}\right)\alpha_j(t)-\sqrt{\kappa}\alpha_{j}^{\rm in}.
\end{equation}
We note that these equations are non-linear because $\delta X$ depends on $\alpha_j$. This non-linear term acts like a displacement-dependent detuning. However, since $\delta X$ is of order one and $g_0\ll\kappa$, this can safely be ignored and we find the steady-state values
\begin{equation}
    \label{eq:avgsstead}
    \alpha_j = -\frac{\sqrt{\kappa}\alpha_j^{\rm in}}{\kappa/2-i\Delta}.
\end{equation}
For the fluctuations, we find the Langevin equations
\begin{equation}
    \label{eq:langevinc1ap2}
    \begin{aligned}
    \partial_t \hat{c}_j(t) = & \left(i\Delta+i\sqrt{2}g_0\delta X-\frac{\kappa}{2}\right)\hat{c}_j(t)-i\sqrt{2}g_j\hat{X}_m(t)\\&-\sqrt{\kappa}\hat{a}_{j}^{\rm in}(t).
    \end{aligned}
\end{equation}
The contribution to the detuning due to the displacement of the mechanical oscillator can again safely be ignored. For the first cavity mode, we directly recover Eq.~\eqref{eq:langevinc1}. For the second cavity mode, we first need to determine the input field operator. 

In the  case of finite coupling efficiency to the cavity mode, the input operator for the first cavity mode reads
\begin{equation}
    \label{eq:inputnoise}
    \hat{a}_1^{\rm in}(t) = \sqrt{\eta_1}\hat{a}_{1}^{\prime\,\rm in}(t)+\sqrt{1-\eta_1}\hat{\nu}_1(t),
\end{equation}
where $\hat{a}_{1}^{\prime\,\rm in}$ denotes the input mode driven by the local oscillator and $\hat{\nu}_1$ denotes an additional source of vacuum noise. Note that $\alpha_1^{\rm in}=\sqrt{\eta_1}\alpha_1'^{\rm in}$, i.e., the cavity mode is driven by a fraction  $\eta_1$ of the physical input power.
The output mode of the first cavity mode that is used for feedback is given by
\begin{equation}
    \label{eq:out1ap}
    \hat{a}^{\prime\,\rm out}_1(t) = \hat{a}_1^{\prime\,\rm in}(t)+\sqrt{\eta_1\kappa}\,\hat{c}_1(t).
\end{equation}
This mode is then combined with an auxiliary local oscillator $\hat{a}_{\rm aux}'$ at a beamsplitter with  splitting ratio $\eta_{\rm aux}$, before being delayed by the time $\tau$. Including further losses arising from the delay  fiber, additional optics as well as coupling to the second cavity mode, the input mode for the second cavity mode reads
{\begin{equation}
    \label{eq:in2ap}
    \begin{aligned}
    &\hat{a}_2^{\rm in}(t) = \sqrt{1-\eta_2\eta_T}\hat{\nu}_2(t) \\&+\sqrt{\eta_2\eta_T}\left[\sqrt{\eta_{\rm aux}}\,e^{i\varphi}\hat{a}_1'^{\rm out}(t-\tau)+\sqrt{1-\eta_{\rm aux}}\,\hat{a}_{\rm aux}'(t-\tau)\right]\\&\hspace{.5cm}
    =\sqrt{1-\eta_2\eta_T}\hat{\nu}_2(t) +\sqrt{\eta_1\eta_2\eta_T\eta_{\rm aux}\kappa}\,e^{i\varphi}\hat{c}_1(t-\tau)\\&+\sqrt{\eta_2\eta_T\eta_{\rm aux}}\,e^{i\varphi}\hat{a}_1'^{\rm in}(t-\tau)+\sqrt{\eta_2\eta_T(1-\eta_{\rm aux})}\,\hat{a}_{\rm aux}'(t-\tau).
    \end{aligned}
\end{equation}
Here the phase $\varphi=\phi_1-\phi_2$ arises because of the different phases that enter the definitions of $\hat{c}_1$ and $\hat{c}_2$, see Eq.~\eqref{eq:displap}. We may now introduce the mode
\begin{equation}
    \label{eq:auxmodeloss}
    \begin{aligned}
    &\hat{a}_{\rm aux}(t) = \sqrt{\frac{\eta_2\eta_T(1-\eta_{\rm aux})}{1-\eta}}\hat{a}'_{\rm aux}(t-\tau)+\sqrt{\frac{1-\eta_2\eta_T}{1-\eta}}\hat{\nu}_2(t)\\&+\sqrt{\frac{(1-\eta_1)\eta_2\eta_T\eta_{\rm aux}}{1-\eta}}\,e^{i\varphi}\left[\sqrt{1-\eta_1}\hat{a}_1'^{\rm in}(t-\tau)-\sqrt{\eta_1}\hat{\nu}_1(t-\tau)\right],
    \end{aligned}
\end{equation}
where we introduced the total efficiency
\begin{equation}
    \label{eq:etatot}
    \eta=\eta_1\eta_2\eta_T\eta_{\rm aux}.
\end{equation}
We note that in the ideal limit where $\eta_1=\eta_2=\eta_T=1$, we have $\hat{a}_{\rm aux}(t)=\hat{a}'_{\rm aux}(t-\tau)$, i.e., we shifted the time-argument. With the help of Eq.~\eqref{eq:auxmodeloss}, we find
\begin{equation}
    \label{eq:a2inap2}
    \hat{a}_2^{\rm in}(t) = \sqrt{\eta}\,e^{i\varphi}\left[\sqrt{\kappa}\hat{c}_1(t-\tau) +\hat{a}_{1}^{\rm in}(t-\tau)\right]+\sqrt{1-\eta}\,\hat{a}_{\rm aux}(t),
\end{equation}
recovering Eq.~\eqref{eq:langevinc2}. Importantly, the mode $\hat{a}_{\rm aux}$ is orthogonal to the mode $\hat{a}_{1}^{\rm in}$ for all values of $\tau$, such that we preserve the commutation relations from the ideal scenario. However, the average of the auxiliary mode reads
\begin{equation}
    \label{eq:auxavap}
    \alpha_{\rm aux} = (1-\eta_1)\sqrt{\frac{\eta_2\eta_T\eta_{\rm aux}}{1-{\eta}}}\alpha_{1}'^{\rm in}+\sqrt{\frac{\eta_2\eta_T(1-\eta_{\rm aux})}{1-\eta}}\alpha_{\rm aux}'.
\end{equation}
For $\eta_1=\eta_2=\eta_T=1$, we obtain $\alpha_{\rm aux}'=\alpha_{\rm aux}$ as expected. However, for $\eta_1\neq1$, $\alpha_{\rm aux}$ is non-zero even in the absence of an auxiliary local oscillator.}

{With these input modes, we find the average displacements of the cavity modes in terms of $\alpha_1^{\rm in} = \sqrt{\eta_1}\,\alpha_1'^{\rm in}$ and $\alpha_{\rm aux}$
\begin{equation}
\label{eq:means}
\begin{aligned}
&\alpha_1 = -\frac{\sqrt{\kappa}\alpha_1^{\rm in}}{\kappa/2-i\Delta},\\& \alpha_2=\frac{1}{\kappa/2-i\Delta}\left[\sqrt{\eta\kappa}\,\frac{{\kappa}/{2}+i\Delta}{{\kappa}/{2}-i\Delta}\alpha_1^{\rm in}-\sqrt{(1-\eta)\kappa}\,\alpha_{\rm aux}\right].
\end{aligned}\vspace{5 pt}
\end{equation}}
The Langevin equations in the main text thus fully capture additional loss channels. However, the {amplitudes of the physical inputs $\alpha_{1}'^{\rm in}$ and $\alpha_{\rm aux}'$ have to be re-scaled in order to take into account all the loss channels that are present}.
\subsection{Frequency space}
\label{app:langfreq}
The Langevin equations in Eqs.~\eqref{eq:langevinmech}, \eqref{eq:langevinc1}, and \eqref{eq:langevinc2} can conveniently be written as a matrix equation in frequency space
\begin{equation}
    \label{eq:langfreq1}
    -i\omega \hat{\boldsymbol{r}}(\omega) = A(\omega)\hat{\boldsymbol{r}}(\omega)+B(\omega) \hat{\boldsymbol{r}}_{\rm in}(\omega),
\end{equation}
where we introduced the vectors
\begin{equation}
\label{eq:vectors}
\hat{\boldsymbol{r}}(\omega)=\colvec{6}{\hat{X}_m(\omega)}{\hat{P}_m(\omega)}{\hat{x}_1(\omega)}{\hat{p}_1(\omega)}{\hat{x}_2(\omega)}{\hat{p}_2(\omega)},\quad\hat{\boldsymbol{r}}_{\rm in}(\omega)=\colvec{5}{\hat{\xi}_{\rm th}(\omega)}{\hat{x}_1^{\rm in}(\omega)}{\hat{p}_1^{\rm in}(\omega)}{\hat{x}_{\rm aux}(\omega)}{\hat{p}_{\rm aux}(\omega)}.
\end{equation}
Here, operators in frequency space are given by
\begin{equation}
\label{eq:fourier}
\hat{O}(\omega) = \frac{1}{\sqrt{2\pi}}\int  \hat{O}(t)\,\,e^{i\omega t}\,dt.
\end{equation}
The matrices in Eq.~\eqref{eq:langfreq1} read

\begin{widetext}
\begin{equation}
\label{eq:matA}
A(\omega) = \begin{pmatrix}
0 & \Omega_{m} & 0 & 0 & 0 & 0 \\
 -\Omega_{m} & -\gamma_m & -2g_1 & 0 & -2g_2 & 0 \\
 0 & 0 & -{\kappa}/{2} & -\Delta & 0 & 0 \\
 -2g_1 & 0 &  \Delta & -{\kappa}/{2} & 0 & 0 \\
  0 & 0 &  -\sqrt{\eta}\kappa\cos(\varphi)\,e^{i\omega\tau} & \sqrt{\eta}\kappa\sin(\varphi)\,e^{i\omega\tau} & -{\kappa}/{2} & -\Delta \\
    -2g_2 & 0 &  -\sqrt{\eta}\kappa\sin(\varphi)\,e^{i\omega\tau} & -\sqrt{\eta}\kappa\cos(\varphi)\,e^{i\omega\tau}  & \Delta & -{\kappa}/{2}
\end{pmatrix},
\end{equation}
and 
\begin{equation}
\label{eq:matB}
B(\omega) = \begin{pmatrix}
0 & 0 & 0 & 0 & 0\\
 -\sqrt{2} & 0 & 0 & 0 & 0\\
 0 & -\sqrt{\kappa} & 0 & 0 & 0\\
 0 & 0 & -\sqrt{\kappa} & 0 & 0 \\
 0 & -\sqrt{\eta\kappa}\cos(\varphi)\,e^{i\omega\tau} & \sqrt{\eta\kappa}\sin(\varphi)\,e^{i\omega\tau}  & -\sqrt{(1-\eta)\kappa} & 0 \\
 0 & -\sqrt{\eta\kappa}\sin(\varphi)\,e^{i\omega\tau} & -\sqrt{\eta\kappa}\cos(\varphi)\,e^{i\omega\tau}  & 0 & -\sqrt{(1-\eta)\kappa} 
\end{pmatrix}.
\end{equation}

From Eq.~\eqref{eq:langfreq1}, we find that any power spectral density can be written using
\begin{equation}
    \label{eq:powspecgen}
    \langle \hat{\boldsymbol{r}}(\omega)\hat{\boldsymbol{r}}^T(\omega')\rangle = C(\omega)\langle \hat{\boldsymbol{r}}_{\rm in}(\omega)\hat{\boldsymbol{r}}_{\rm in}^T(\omega')\rangle C^T(\omega'),
\end{equation}
where $C(\omega)=[A(\omega)+i\omega]^{-1}B(\omega)$ and the input spectral density matrix reads
\begin{equation}
\label{eq:matin}
\langle \hat{\boldsymbol{r}}_{\rm in}(\omega)\hat{\boldsymbol{r}}_{\rm in}^T(\omega')\rangle = \delta(\omega+\omega')\begin{pmatrix}

S_{\rm th}(\omega) & 0 & 0 & 0 & 0\\
 0 & 1/2 & i/2 & 0 & 0\\
 0 & -i/2 & 1/2 & 0 & 0 \\
 0 &0  & 0  & 1/2 & i/2 \\
 0 & 0 & 0 & -i/2 & 1/2
\end{pmatrix},
\end{equation}
with $S_{\rm th}(\omega)$ given in Eq.~\eqref{eq:thermalspec}.

{
Before solving these equations, it is instructive to consider the simplified scenario where $\Delta=0$ and $\varphi=\pi/2$. From the equation for the first cavity mode, we find
\begin{equation}
    \label{eq:freqp1}
    \left(\frac{\kappa}{2}-i\omega\right)\hat{p}_1(\omega) = -2g_1\hat{X}_m(\omega)-\sqrt{\kappa}\hat{p}_1^{\rm in}(\omega),
\end{equation}
illustrating how the momentum quadrature contains information on the position of the mechanical resonator; $\hat{p}_1^{\rm in}$ denotes the input (vacuum) noise.
The momentum quadrature of the first cavity mode is then fed to the second cavity mode and we find
\begin{equation}
 \left(\frac{\kappa}{2}-i\omega\right)\hat{x}_2(\omega) = \sqrt{\eta\kappa}e^{i\omega\tau}\left[\sqrt{\kappa}\hat{p}_1(\omega)+\hat{p}_1^{\rm in}\right]-\sqrt{(1-\eta)\kappa}\hat{x}_{\rm aux}.
\end{equation}}

{
Finally, the position quadrature of the second cavity mode couples to the mechanical oscillator, completing the feedback loop. We find for the momentum quadrature of the mechanical oscillator
\begin{equation}
\label{eq:pmfreq}
    -i\omega \hat{P}_m(\omega) = -\Omega_m\hat{X}_m(\omega)-\gamma_m\hat{P}_m(\omega)-4\, e^{i\omega\tau}\frac{g_1g_2\sqrt{\eta}\kappa}{\left({\kappa}/{2}-i\omega\right)^2}\hat{X}_m(\omega)+\sqrt{2}\hat{\xi}_{\rm th}(\omega)+\sqrt{2}\hat{\xi}_{\rm fb}(\omega).
\end{equation}
Using the relation $-i\omega\hat{X}_m(\omega)=\Omega_m\hat{P}_m(\omega)$, we find that the real and imaginary parts of the pre-factor of $\hat{X}_m(\omega)$ in Eq.~\eqref{eq:pmfreq} correspond to frequency shift and damping respectively. Since the delay $\tau$ determines the phase of the pre-factor, it can be used to tune between a frequency shift and damping, in complete agreement with our analysis in the main text. Finally, to recover Eqs.~\eqref{eq:dampshift} in the main text, we neglect the frequency dependence of the frequency shift and damping and take the unresolved sideband limit $\kappa\gg\Omega_m$.}

{\subsection{Eliminating the cavity}}
Since we are dealing with a linear set of equations, the cavity modes may be eliminated from Eq.~\eqref{eq:langfreq1}. A tedious but straightforward calculation results in
\begin{equation}
\label{eq:pred}
-i\omega \hat{P}_m(\omega) = -[\Omega_{m}+2\delta\Omega_m(\omega)]\hat{X}_m(\omega)-[\Gamma_m(\omega)+\gamma_m]\hat{P}_m(\omega)+\sqrt{2}\hat{\xi}_{\rm th}(\omega)+\sqrt{2}\hat{\xi}_{\rm fb}(\omega),
\end{equation}
where we introduced the frequency-dependent frequency shift
\begin{equation}
\label{eq:delomom}
\delta\Omega_m(\omega)={\rm Re}\left\{\frac{2\Delta(g_1^2+g_2^2)}{\Delta^2+\left({\kappa}/{2}-i\omega\right)^2}-2\,\,e^{i\omega\tau}g_1g_2\sqrt{\eta}\kappa\frac{2\Delta\left({\kappa}/{2}-i\omega\right)\cos(\varphi)-\left[\Delta^2-\left({\kappa}/{2}-i\omega\right)^2\right]\sin(\varphi)}{\left[\Delta^2+\left({\kappa}/{2}-i\omega\right)^2\right]^2}\right\},
\end{equation}
and the optomechanical damping rate
\begin{equation}
\label{eq:gammaom}
\Gamma_m(\omega)=\frac{\Omega_{m}}{\omega}{\rm Im}\left\{-\frac{4\Delta(g_1^2+g_2^2)}{\Delta^2+\left({\kappa}/{2}-i\omega\right)^2}+4\,e^{i\omega\tau}g_1g_2\sqrt{\eta}\kappa\frac{2\Delta\left({\kappa}/{2}-i\omega\right)\cos(\varphi)-\left[\Delta^2-\left({\kappa}/{2}-i\omega\right)^2\right]\sin(\varphi)}{\left[\Delta^2+\left({\kappa}/{2}-i\omega\right)^2\right]^2}\right\}.
\end{equation}
The feedback noise is described by the spectral density {$\langle\hat{\xi}_{\rm fb}(\omega)\hat{\xi}_{\rm fb}(\omega')\rangle=\delta(\omega+\omega')\,S_{\rm fb}(\omega)$ with 
\begin{equation}
\label{eq:fbnoise}
S_{\rm fb}(\omega)=\frac{\kappa(g_1^2+g_2^2)}{(\kappa/2)^2+(\Delta+\omega)^2}+2\kappa g_1g_2\sqrt{\eta}\frac{[(\Delta+\omega)^2-(\kappa/2)^2]\cos(\varphi+\omega\tau)+\kappa(\Delta+\omega)\sin(\varphi+\omega\tau)}{[(\kappa/2)^2+(\Delta+\omega)^2]^2}.
\end{equation}}
We stress that Eqs.~(\ref{eq:pred}-\ref{eq:fbnoise}) involve no further approximations after the linearization of the optomechanical coupling.
\end{widetext}

For a high-quality oscillator, we may replace $\delta\Omega_m(\omega)$ with $\delta\Omega_m\equiv \delta\Omega_m(\Omega_m)$ and $\Gamma_m(\omega)$ with $\Gamma_m$ in Eq.~\eqref{eq:pred}. This results in a second-order differential equation describing a damped harmonic oscillator
\begin{equation}
\label{eq:dampharmx2}
\begin{aligned}
\partial_t^2\hat{X}_m(t) =& -(\Omega_{m}+\delta\Omega_m)^2\hat{X}_m(t)-(\Gamma_m+\gamma_m)\,\partial_t\hat{X}_m(t)\\&+\Omega_{m}\sqrt{2}\,\hat{\xi}_{\rm th}(t)+\Omega_{m}\sqrt{2}\,\hat{\xi}_{\rm fb}(t).
\end{aligned}
\end{equation}
From this equation, we may derive the spectral density {
\begin{equation}
\label{eq:specdensx}
\begin{aligned}
S_{XX}(\omega)=&\frac{1}{2}\frac{S_{\rm th}(\Omega_{m})+S_{\rm fb}(\Omega_{m})}{(\Omega_{m}+\delta\Omega_m-\omega)^2+[({\gamma_m+\Gamma_m})/{2}]^2}\\&+\frac{1}{2}\frac{S_{\rm th}(-\Omega_{m})+S_{\rm fb}(-\Omega_{m})}{(\Omega_{m}+\delta\Omega_m+\omega)^2+[({\gamma_m+\Gamma_m})/{2}]^2},
\end{aligned}
\end{equation}}
where we again invoked the high quality factor of the oscillator.

The number of phonons can be obtained from
\begin{equation}
\label{eq:nphon}
\begin{aligned}
2\bar{n}_m+1 = &\int_{-\infty}^{\infty}[S_{XX}(\omega)+S_{PP}(\omega)]\,\frac{d\omega}{2\pi}\\&=\int_{-\infty}^{\infty}S_{XX}(\omega)[1+(\omega/\Omega_{m})^2]\,\frac{d\omega}{2\pi}\\&\simeq 2\int_{-\infty}^{\infty}S_{XX}(\omega)\,\frac{d\omega}{2\pi},
\end{aligned}
\end{equation}
which yields Eq.~\eqref{eq:number} {upon using $\Gamma_m={S}_{\rm fb}(\Omega_m)-{S}_{\rm fb}(-\Omega_m)$}.

\subsection{The unresolved sideband limit}
\label{ap:theo_unr}
Here we provide simplified expressions for the unresolved sideband limit, $\Omega_m\ll\kappa$. In contrast to the expressions given in the main text, we consider a finite detuning $\Delta$. We first consider dynamical backaction cooling without coherent feedback by setting $\eta=0$. In this case, the frequency shift and the optomechanical damping rate reduce to
\begin{equation}
\label{eq:domedoppeta0}
\delta\Omega_{\rm dyn} = \frac{2\Delta(g_1^2+g_2^2)}{\Delta^2+\left({\kappa}/{2}\right)^2},
\end{equation}
and
\begin{equation}
\label{eq:gammadoppeta0}
\Gamma_{\rm dyn}=-4\frac{\Delta\kappa\Omega_{m}(g_1^2+g_2^2)}{\left[\Delta^2+\left({\kappa}/{2}\right)^2\right]^2}.
\end{equation}
For $\Gamma_m\gg\gamma_m$, the phonon occupation number in Eq.~\eqref{eq:number} reduces to
\begin{equation}
\label{eq:nmindoppeta0}
\bar{n}_{\rm dyn}=\frac{\Delta^2+\left({\kappa}/{2}\right)^2}{4|\Delta|\Omega_{m}}-\frac{1}{2},
\end{equation}
which is minimized at $\Delta=-\kappa/2$ where it reads
\begin{equation}
\label{eq:nmindoppeta02}
\bar{n}_{\rm dyn}=\frac{\kappa}{4\Omega_{m}}\gg 1.
\end{equation}
Here, we dropped the term $1/2$ as it becomes negligible. The last equation is the well-known  cooling limit for cavity \textit{dynamical backaction cooling} in the unresolved sideband regime \cite{aspelmeyer2014}.

In the presence of coherent feedback ($\eta\neq0$), we find the following expressions to lowest-order in $\Omega_{m}/\kappa$
\begin{widetext}
\begin{equation}
\label{eq:domedopp}
\delta\Omega_m = \frac{2\Delta(g_1^2+g_2^2)}{\Delta^2+\left({\kappa}/{2}\right)^2}-2\cos(\Omega_{m}\tau)g_1g_2\sqrt{\eta}\kappa\frac{\Delta\kappa\cos(\varphi)-\left[\Delta^2-\left({\kappa}/{2}\right)^2\right]\sin(\varphi)}{\left[\Delta^2+\left({\kappa}/{2}\right)^2\right]^2},
\end{equation}
and
\begin{equation}
\label{eq:gammadopp}
\Gamma_m=4\sin(\Omega_{m}\tau)g_1g_2\sqrt{\eta}\kappa\frac{\Delta\kappa\cos(\varphi)-\left[\Delta^2-\left({\kappa}/{2}\right)^2\right]\sin(\varphi)}{\left[\Delta^2+\left({\kappa}/{2}\right)^2\right]^2}.
\end{equation}
We note that the coherent feedback allows for a $\Gamma_m$ that is independent of $\Omega_m/\kappa$ to lowest-order in this parameter. This stands in contrast to cavity dynamical backaction cooling, where $\Gamma_{\rm dyn}\propto\Omega_m/\kappa $ to lowest-order [cf.~Eq.~\eqref{eq:gammadoppeta0}]. In the limit $\Gamma_m\gg\gamma_m$, the number of phonons is then given by
\begin{equation}
    \label{eq:nbadcav}
    \bar{n}_m = \frac{\kappa}{\Gamma_m}\frac{(g_1^2+g_2^2)}{(\kappa/2)^2+\Delta^2}+\frac{1}{2}\frac{\cos(\Omega_{m}\tau)}{\sin(\Omega_{m}\tau)}\frac{\kappa\Delta\sin(\varphi)+[\Delta^2-(\kappa/2)^2]\cos(\varphi)}{\kappa\Delta\cos(\varphi)-[\Delta^2-(\kappa/2)^2]\sin(\varphi)}-\frac{1}{2}.
\end{equation}
For $\Delta=0$, Eqs.~\eqref{eq:domedopp}, \eqref{eq:gammadopp}, and \eqref{eq:nbadcav} reduce to Eqs.~\eqref{eq:dampshift} and \eqref{eq:numberlim}.

\end{widetext}

\section{Equivalence with measurement-based feedback cooling} 
\label{ap:measfb}

In this section, we extend the comparison between coherent feedback and measurement-based feedback of Sec.~\ref{sec:meascool} by considering a generic spectral filter function $\Xi_\mathrm{mf}(\omega)$, which implements an arbitrary feedback response taking place after the measurement. As for the treatment of coherent feedback, we can obtain a reduced description for the mechanical variables by eliminating the cavity mode 
\begin{widetext}
\begin{equation}
\label{eq:Pmf}
-i\omega \hat{P}_m(\omega) = -[\Omega_{m}+2\delta\Omega_\mathrm{mf}(\omega)]\hat{X}_m(\omega)-[\Gamma_\mathrm{mf}(\omega)+\gamma_m]\hat{P}_m(\omega)+\sqrt{2}\,\hat{\xi}_{\rm th}(\omega)+\sqrt{2}\,\hat{\xi}_\mathrm{mf}(\omega),
\end{equation}
where we introduced the frequency shift and optomechanical damping rate, respectively given by
\begin{align}
\delta\Omega_\mathrm{mf}(\omega)&=\frac12 {\rm Re}\left\{\frac{2g_1\Xi_\mathrm{mf}(\omega)}{{\kappa}/{2}-i\omega} \right\} \,, \label{eq:Omega_mf} \\
\Gamma_\mathrm{mf}(\omega)&=-\frac{\Omega_{m}}{\omega} {\rm Im}\left\{\frac{2g_1\Xi_\mathrm{mf}(\omega)}{{\kappa}/{2}-i\omega} \right\}  \,, \label{eq:Gamma_mf}
\end{align}
 with $\Gamma_\mathrm{mf}=\Gamma_\mathrm{mf}(\Omega_m)$, and collected the noise terms in the expression {
\begin{equation}\label{eq:filtermf}
\hat{\xi}_\mathrm{mf}(\omega)= \frac{g_1\sqrt{2\kappa}}{{\kappa}/{2}-i\omega} \hat{x}_\mathrm{in}(\omega) -\frac{\Xi_\mathrm{mf}(\omega)}{\sqrt{2\kappa}}\left(\frac{{\kappa}/{2}+i\omega}{{\kappa}/{2}-i\omega}\right) \hat{p}_\mathrm{in}(\omega)- \frac{\Xi_\mathrm{mf}(\omega)}{\sqrt{2\kappa}}\sqrt{\frac{1-\eta_\mathrm{det}}{\eta_\mathrm{det}}}\hat{p}_0(\omega) \,.
\end{equation}}
\end{widetext}
The first term in Eq.~\eqref{eq:filtermf} describes shot noise due to radiation pressure interaction with the first cavity, while the second and third terms describe the noise associated with the feedback process. The spectral density characterizing the noise $\hat{\xi}_\mathrm{mf}$ reads

{
\begin{equation}
\begin{aligned} \label{eq:S_mf}
S_\mathrm{mf}(\omega)&= \frac{\kappa g_1^2}{\omega^2+\left({\kappa}/{2}\right)^2} + \frac{\vert \Xi_\mathrm{mf}(\omega)\vert^2}{4\kappa \eta_\mathrm{det}}- g_1{\rm Im}\left\{\frac{\Xi_\mathrm{mf}(\omega)}{{\kappa}/{2}-i\omega} \right\} \\
&=\frac{\kappa g_1^2}{\omega^2+\left({\kappa}/{2}\right)^2} + \frac{\vert \Xi_\mathrm{mf}(\omega)\vert^2}{4\kappa \eta_\mathrm{det}}+ \frac12 \frac{\omega}{\Omega_{m}}\Gamma_\mathrm{mf}(\omega) \,.
\end{aligned}
\end{equation}}
We can now compare the quantities $\delta\Omega_\mathrm{mf}(\omega)$, $\Gamma_\mathrm{mf}(\omega)$ and $S_\mathrm{mf}(\omega)$ with the corresponding expressions  derived for the case of coherent feedback, respectively given by Eqs.~\eqref{eq:delomom},~\eqref{eq:gammaom} and~\eqref{eq:fbnoise}. We further focus on the case $\Delta=0$ and $\varphi=\pi/2$, which is relevant for cooling.

By requiring the measurement-based feedback to induce the same effective damping and broadening as coherent feedback at all frequencies, i.e., by enforcing $\delta\Omega_\mathrm{mf}(\omega)=\delta\Omega_m(\omega)$ and $\Gamma_\mathrm{mf}(\omega)=\Gamma_m(\omega)$,  we get the following filter function
 \begin{equation}\label{eq:Ximeas}
\widetilde{\Xi}_\mathrm{mf}(\omega)=-4g_2 \sqrt{\eta}\frac{\,e^{i\omega\tau}}{1-2i{\omega}/{\kappa}}\,.
 \end{equation}

In the above expression, the exponential term describes the effect of delay, i.e., in order to reproduce the effect of coherent feedback a delay has to be incorporated in the measurement-based feedback loop. If we compare this expression to the case of cold damping in Eq.~\eqref{eq:Hfb}, we notice that $\kappa/2$ plays the role of the feedback bandwidth, $4g_2 \sqrt{\eta}$ the strength of the feedback and that there is no frequency dependence in the numerator for $\tau=0$. 

We then plug this filter in Eq.~\eqref{eq:S_mf} and compare the corresponding noise spectral density with that of coherent feedback Eq.~\eqref{eq:fbnoise}. 
It is straightforward to show that whenever $\eta=\eta_\mathrm{det}$, the ensuing noise spectral densities $S_\mathrm{fb}(\omega)$ and $S_\mathrm{mf}(\omega)$ match for all frequency values and for arbitrary values of delay. This shows the equivalence of measurement-based feedback cooling and coherent feedback cooling in more general terms than the particular case of cold damping. 

{
\section{Coherent feedback cooling performance in different cavity regimes}
\label{ap:coolingperformance}
Here, we investigate the comparative performance of coherent feedback cooling versus standard cavity dynamical backaction cooling in both the resolved and unresolved sideband regimes. To quantify performance, we examine the lowest achievable phonon number under different scenarios: For cavity cooling, we consider the well-known cases of $\Delta=-\Omega_m$ for $\kappa\ll\Omega_m$ and $\Delta=-\kappa/2$ for $\kappa\gg\Omega_m$, where the minimum phonon number asymptotically approaches  $\kappa/(4\Omega_m)$. In Fig.~\ref{fig:sidebandpar_eta} we compare these cooling strategies to coherent feedback cooling, for $\Delta=-\Omega_m$ and $\Delta=0$, which are the optimal detunings in the two different cavity regimes. Our results demonstrate that, in the resolved sideband regime, coherent feedback cooling provides no significant advantage over cavity cooling when operating at $\Delta=-\Omega_m$ and using the optimal parameters $\varphi=\Omega_m\tau=\pi/2$ mentioned in the main text. However, in the unresolved sideband regime, we find that coherent feedback cooling (which performs best for $\Delta = 0$)  outperforms standard optomechanical cooling (which leads to minimal occupation at $\Delta=-\kappa/2$). Intriguingly, we observe that coherent feedback cooling performs better deep in the unresolved sideband regime, where the cavity can actually be ignored and the first interaction is performing an almost instantaneous readout of the membrane's motion. The plot is shown for $\eta\simeq 0.98$. For more realistic scenarios, the minimal phonon number becomes independent of $\kappa$ early in the unresolved sideband regime, saturating to the value expressed in the main text in Eq.~\eqref{eq:numberlim}. 
    \begin{figure}[ht]
    \includegraphics[scale=1]{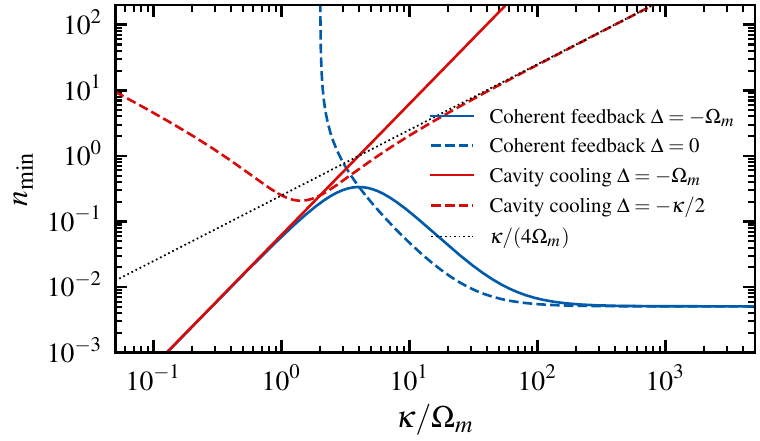}
    \caption{ {Cooling performance for different cavity regimes for coherent feedback and cavity dynamical backaction cooling. Blue lines show the performance for coherent feedback for $\Delta=-\Omega_m$ (solid) and $\Delta=0$ (dashed), in both cases $\varphi=\Omega_m\tau=\pi/2$, as described in the theory section of the manuscript. Red lines show the performance of cavity cooling for $\Delta=-\Omega_m$ (solid) and $\Delta=-\kappa/2$ (dashed). The black dotted line shows the cavity cooling limit in the unresolved sideband regime. The coherent feedback curves are shown for $\eta\simeq 0.98$. }}
    \label{fig:sidebandpar_eta}
    \end{figure}   
}

\newpage
\bibliography{CoherentFeedback}

\end{document}